\documentclass[preprint2]{aastex}
\usepackage{natbib}
\def\doeps{1}	% =1 to include the epsf figures or, with 0:
\def\dovsp{0}	% =1 for blank space, =0 for captions only
\ifnum\doeps=1\input epsf\else\fi
\newcommand{\FIG}[3]{\begin{figure}[!th]
\ifnum\doeps=0\ifnum\dovsp=1\vspace{#2}\else\vspace{24pt}\fi
\else\centerline{\epsfxsize=#2\epsffile{#1.ps}}\fi\par
\caption[]{#3\label{#1}}
\end{figure}}
\newcommand{\EQ}{\begin{equation}}
\newcommand{\EN}{\end{equation}}
\newcommand{\EQA}{\begin{eqnarray}}
\newcommand{\ENA}{\end{eqnarray}}
\newcommand{\UQ}{\begin{displaymath}}
\newcommand{\UN}{\end{displaymath}}

\newcommand{\NOCOMMENT}[1]{}

\begin{document}
%%%%%%%%%%%%%%%%%%%%%%%%%%%%%%%%%%%%%%%%%%%%%%%%%%%%%%%%%%%%%%%%%%%%

\title{Solar Oscillations and Convection: II. Excitation of Radial 
Oscillations}

\author{R. F. Stein}
\affil{Dept. of Physics and Astronomy, Michigan State University,
East Lansing, MI 48823, U.S.A.}
\email{stein@pa.msu.edu}

\and 

\author{\AA. Nordlund}
\affil{Theoretical Astrophysics Center and 
       Astronomical Observatory / NBIfAFG, \\
       Juliane Maries Vej 30, Dk-2100 Copenhagen \O, Denmark}
\email{aake@astro.ku.dk}

%%%%%%%%%%%%%%%%%%%%%%%%%%%%%%%%%%%%%%%%%%%%%%%%%%%%%%%%%%%%%%%%%%%%
\begin{abstract}

Solar $p$-mode oscillations are excited by the work of stochastic,
non-adiabatic, pressure fluctuations on the compressive modes.  We
evaluate the expression for the radial mode excitation rate derived by
\citet{Nordlund+Stein98form} using numerical simulations of near
surface solar convection.  We first apply this expression to the three
radial modes of the simulation and obtain good agreement between the
predicted excitation rate and the actual mode damping rates as determined from
their energies and the widths of their resolved spectral profiles.
These radial simulation modes are essentially the same as the solar
modes at the resonant frequencies, where the solar modes have a node at
the depth of the bottom of the simulation domain.  We then apply this
expression for the mode excitation rate to the solar modes and obtain
excellent agreement with the low $\ell$ damping rates determined from
GOLF data.  Excitation occurs close to the surface, mainly in the
intergranular lanes and near the boundaries of granules (where
turbulence and radiative cooling are large).  The non-adiabatic pressure
fluctuations near the surface are produced by small instantaneous local
imbalances between the divergence of the radiative and convective
fluxes near the solar surface.  Below the surface, the non-adiabatic
pressure fluctuations are produced primarily by turbulent pressure
fluctuations (Reynolds stresses).  The frequency dependence of the mode
excitation is due to effects of the mode structure and the pressure
fluctuation spectrum.  Excitation is small at low frequencies due to
mode properties -- the mode compression decreases and the mode mass
increases at low frequency.  Excitation is small at high frequencies
due to the pressure fluctuation spectrum -- pressure fluctuations
become small at high frequencies because they are due to convection
which is a long time scale phenomena compared to the dominant $p$-mode
periods.

\end{abstract}


%%%%%%%%%%%%%%%%%%%%%%%%%%%%%%%%%%%%%%%%%%%%%%%%%%%%%%%%%%%%%%%%%%%%

\section{Introduction}

Two ideas for the source of $p$-mode excitation have been pursued -- 
overstability (as in pulsating stars) and turbulent Reynolds stresses 
(as in jet noise) 
\citep{Biermann48,Schwarzschild48, %
Lighthill52, %
Stein67,Stein68, %
Crighton75, %
Ando+Osaki77,Goldreich+Keeley77b, %
Goldreich+Kumar90, %
Balmforth92c, %
Musielak+94waves}.
We show here that it is the $PdV$ work of stochastic, non-adiabatic 
pressure fluctuations which is the primary mode excitation mechanism 
\citep{Stein+Nordlund91osc,Bogdan+93,Goldreich+94b}.
Near the surface, the non-adiabatic gas pressure (i.e. entropy) fluctuations
dominate.  They are produced by radiative cooling at the solar surface
which is not locally and instantaneously exactly balanced by convective
heat deposition.  Below the surface, non-adiabatic turbulent pressure
(Reynolds stress) fluctuations dominate.  They are produced by the
turbulent convective motions.

In a previous paper \citep{Nordlund+Stein98form} we derived an exact
expression for the stochastic excitation rate of the radial solar 
$p$-modes by the $PdV$ work of 
non-adiabatic gas and turbulent pressure fluctuations on the mode
compression.  We now use realistic numerical simulations of near surface
solar convection (depth about 2.5 Mm) to evaluate this expression 
\citep{Stein+Nordlund98gran}.
Because the largest entropy and pressure fluctuations occur near the
solar surface and because modes with frequencies in the 3 -- 4 mHz range,
where the excitation rate is largest, are confined near the solar
surface, these near surface simulations include the primary excitation
and damping processes.  

\section{Mode Excitation: Formalism}

The rate of energy input to the modes can be calculated starting with
the kinetic energy equation for the modes
\citep{Nordlund+Stein98form}.  Neglecting the viscous stresses,
\begin{eqnarray}
\rho {D \over {D t}} \left( {1 \over 2} u_z^2 \right) & = &
- {\partial \over {\partial z}} \left[ u_z \left( P_g + P_t \right)
\right] + \rho u_z g \nonumber\\
& + & \left( P_g + P_t \right) {{\partial u_z} \over {\partial z}}
\ .
\end{eqnarray}
After integrating this equation over depth, the flux divergence term
only gives contributions from the end points and is negligible.  The
buoyancy term is small because mass is conserved so there is no net
mass flux.  The last term is the $PdV$ work,
\begin{equation}
W = \int dt \int dz \delta P {{\partial \dot{\xi}} \over
    {\partial z}}
\ .
\end{equation}
There are several contributions to this work.  
The displacement, $\xi$,  has contributions from the modes as well as
the random convective motions.  The pressure, $\delta P$, has coherent
contributions from the
modes and incoherent contributions from the random convective motions.
Both coherent and incoherent contributions can be further divided into
adiabatic and non-adiabatic terms.  The dominant driving comes from the
interaction of the non-adiabatic, incoherent pressure fluctuations,
\begin{equation}
\delta P^{\rm nad} = \left( \delta \ln P - \Gamma_1 \delta \ln \rho \right)
P \ ,
\end{equation}
with the coherent mode displacement,
\begin{equation}
{{D \ln \rho} \over {D t}} = - {{\partial \dot{\xi}} \over {\partial z}}
\ .
\end{equation}
This is a stochastic process, so the pressure fluctuations occur with
random phases with respect to the modes.  Therefore one must average over
all possible relative phases between them.  The resulting rate of energy
input to the modes is \citep{Nordlund+Stein98form}
\begin{equation}
{{\Delta \langle E_{\omega} \rangle} \over {\Delta t}} =
{{\omega^2 \left | \int_r {d r \delta P^*_{\omega} {{\partial
\xi_{\omega}}
 \over {\partial z}}} \right |^2} \over {8 \Delta \nu E_{\omega}}}
\ .
\label{eq_edot}
\end{equation}
Here, $\delta P_{\omega}$ is the time Fourier transform of the 
non-adiabatic total pressure.  $\Delta \nu =$ 1/(total time interval)
is the frequency resolution with which $\delta P_{\omega}$ is computed.
$\xi_{\omega}$ is the mode displacement eigenfunction, which is typically
chosen to be real for an adiabatic mode.  In that case, taking the 
complex conjugate of the pressure is not necessary, but we retain it
for generality.  The mode energy is
\begin{equation}
E_{\omega} = {1 \over 2} \omega^2 \int_{r} dr ~ \rho ~ \xi_{\omega}^2
\left({r \over R}\right)^2 = M_{\omega} V^2_{\omega}(R)
\ .
\label{eq_enorm}
\end{equation}
Here $M_{\omega}$ is the mode mass and $V_{\omega}(R)$ is the mode
velocity amplitude at the surface.
Eqn~\ref{eq_edot} is similar to the expression of \citet{Goldreich+94b}
(eqn. 26), but involves no approximations.
Having the numerical simulation data, we can evaluate this expression
exactly without having to make approximations in order to evaluate it
analytically \footnote{A detailed description of how eqn~\ref{eq_edot} 
is evaluated is given in the appendix.  Its validity is tested by
applying it to the three radial modes of the simulation domain 
(sec.~\ref{sec:excite_sim}).}

\section{Simulations}

We simulate a small portion of the solar photosphere and the upper
layers of the convection zone, a region extending $6 \times 6$ Mm
horizontally and from the temperature minimum at $-0.5$ Mm down to
$2.59$ Mm below the visible surface.  We solve the equations of mass,
momentum and energy conservation in the form:\\
\EQA
    { { \partial \ln \rho } \over { \partial t } } &=&
    - {\bf u} \cdot \nabla \ln \rho
    - \nabla \cdot {\bf u}
\ , \label{mass.eq} \\
    { { \partial {\bf u} } \over { \partial t } } &=&
    - {\bf u} \cdot \nabla {\bf u}
    + {\bf g}
    - { P \over \rho } \nabla \ln P  
    + {1 \over {\rho}} \nabla \cdot \underline{\underline{\sigma}}
\ , \label{mom.eq} \\
    { { \partial e } \over { \partial t } } &=&
    - {\bf u} \cdot \nabla e
    - { P \over \rho } \nabla \cdot {\bf u}
\nonumber \\ &&
    + Q_{rad} + Q_{visc}
    \ , \label{energy.eq}
\ENA
where $\underline{\underline{\sigma}}$ is the viscous stress tensor,
$Q_{rad}$ is the radiative heating and $Q_{visc}$ is the viscous
dissipation.

We use a non-staggered grid of either $125^2$ cells horizontally 
$\times 82$ cells vertically or $63^3$ cells.  The
spatial derivatives are calculated using third order splines and the
time advance is a third order leapfrog scheme
\citep{Hyman1979,Nordlund+Stein90a}.  The code is stabilized by a
hyper-viscosity which removes short wavelength noise without damping
the longer wavelengths.

A large fraction of the internal energy is in the form of ionization
energy near the solar surface, so we use a realistic equation of state
(including the effects of ionization and excitation of hydrogen and
other abundant elements and the formation and ionization of $H_{2}$ 
molecules).  The
pressure is found by interpolation in a table of $P(\ln \rho, e)$ and
its derivatives which is calculated with the Uppsala stellar atmosphere
package \citep{Gustafsson+75}.  

Radiative energy exchange is critical in determining the structure of
the upper convection zone.  Near the surface of the sun, the energy
flow changes from almost exclusively convective below the surface to
radiative above the surface.  The interaction between convection and
radiation near the solar surface determines what we observe and
produces the entropy fluctuations that lead to the buoyancy work which
drives the convection and that give rise to the non-adiabatic pressure
fluctuations which excite the p-mode oscillations.
Hence, the interaction between convection and
radiation is crucial for both the diagnostics and the dynamics of
convection.   Since the top of the convection zone occurs near the
level where the continuum optical depth is one, neither the optically
thin nor the diffusion approximations give reasonable results.  We
therefore include 3D, LTE, non-gray radiation transfer in our model.

We simulate only a small region near the solar surface and must
therefore impose boundary conditions inside the convectively unstable
region.  What happens outside our computational domain in principle
influences the convective motions inside.  However, at the top
boundary, the density is very low relative to the rest of the volume
and hence whatever happens there has very little influence on the
convective motions.  At the bottom, the incoming fluid is to a very
good approximation isentropic and featureless, and hence carries little
information.  The unknown influence of external regions should
therefore be small.  This assertion is indeed confirmed by experiments
with boundaries located at different depths.

The horizontal directions are taken to be periodic.  In the vertical
direction, we have a transmitting boundary at the temperature minimum
\citep{Nordlund+Stein90a}.  This is achieved by a larger than normal
zone at the top boundary.  Across this zone we make the vertical
derivative of the density hydrostatic, set the vertical derivative of
the velocity zero and hold the internal energy at the top fiducial
layer constant in time and space.  At the bottom of the computational
domain, outgoing fluid goes out with whatever properties it has.  For
incoming fluid, we adjust the pressure such that the net mass flux
through the bottom boundary vanishes.  (This ensures that there is no
boundary work done on vertical oscillation modes.)  The pressure on the
bottom boundary thus varies in time but is uniform over the horizontal
plane.  We damp fluctuations of the horizontal and vertical velocity of
the incoming fluid, using a long time constant. Finally, we adjust the
density and energy of the incoming fluid, at constant pressure, to fix
its entropy (in both space and time).

The ability to do a direct numerical simulation with the wide range of 
length scales 
matching the dimensionless parameters -- Reynolds number, Rayleigh number
and Prandtl number -- of the solar convection zone is beyond the speed
and memory capabilities of current computers.  Thus, our simulations
are of the type called ``Large Eddy Simulations".  It is, however,
possible to resolve the surface thermal boundary layer of the
convection zone and this we have done.  Indeed, this is required to
achieve results that agree quantitatively with solar observations
\citep{Stein+Nordlund2000soho}.

The picture of convection that has emerged from the simulations
is the following:  Convection is driven by radiative cooling in the
thin thermal boundary layer at the solar surface.  It consists of cool,
low entropy, filamentary, turbulent, downdrafts that plunge through a
warm, entropy neutral, smooth, diverging, laminar upflow.  Upflows must
diverge as they ascend into lower density layers in order to conserve
mass.  This divergence smooths out any variations in their properties
that might arise.  Only a small fraction of the fluid at depth reaches
the surface to be cooled and form the cores of the downdrafts.  Most
fluid turns over within a scale height and is entrained by the
downdrafts.  These low entropy downflows are the site of most of the
buoyancy work that drives the convection.  (See \citet{Stein+Nordlund98gran}
for more details.)

We have made numerous comparisons between the predictions of
the simulations and solar observations:
\begin{itemize}
\itemsep=1pt
\parskip=1pt
\item[$\bullet$] The depth of the convection zone depends on the
opacities that determine the temperature versus pressure (and hence
entropy) stratification of the surface layers, the spectral line
blocking, the convective
efficiency of the superadiabatic layers immediately below the surface
that determines the transition to the asymptotic adiabat, and the
equation of state that determines the further run of temperature versus
pressure through the convection zone.  Excellent agreement is obtained
between the depth of the convection zone predicted by our numerical
simulations and that inferred from helioseismology
\citep{Rosenthal+98freq}.

\item[$\bullet$] The cavity for high frequency modes is enlarged by
turbulent pressure support and 3D non-linear opacity effects which
increase the average temperature required to produce a given effective
temperature in an inhomogeneous compared to a homogeneous atmosphere.
The $p$-mode eigenfrequencies calculated from the mean simulation
atmosphere are significantly closer to the observed mode frequencies
than those for standard spherically symmetric, mixing length models
\citep{Rosenthal+98soho,Rosenthal+98freq}.

\item[$\bullet$] The simulation granulation size spectrum and the
distribution of emergent intensities, when smoothed by the point spread
function appropriate for the Swedish Vacuum Solar Telescope on La Palma
(which produces the best solar images available today), closely match
the observations \citep{Stein+Nordlund98gran}.

\item[$\bullet$] The width of photospheric iron lines (whose thermal
speed is small) is a signature of the convective velocities.  The net
Doppler blue shift and asymmetry of spectral lines is a signature of
the correlation between velocity and temperature fluctuations.  Both
these signatures agree closely with observations \citep{Asplund+99b}.

\end{itemize}
This gives us confidence that the simulations are properly modeling
the crucial properties of near surface solar convection.

\section{Mode Excitation: Simulation} \label{sec:excite_sim}

%Simulation results
Three radial modes exist in our simulation and we first apply
eqn~(\ref{eq_edot}) to these modes and compare the rate of work
on the modes it predicts with the actual excitation rate 
$dE_{\omega}/dt = \Gamma E_{\omega}$ determined from the mode's 
energies and widths in the simulation.
The modes can be clearly
seen in the spectrum of the horizontally averaged, depth integrated
kinetic energy (fig~\ref{figure1}).  Their properties are
given in table 1.
The lowest mode (with no zero crossings inside the computational 
domain) has a frequency of 2.57 mHz and a FWHM of 19 $\mu$Hz.
Hence a simulation significantly longer than 14.5 hrs is required to
resolve this mode.  We use a simulation of 43 hours, with a spatial
resolution of $63^3$.  Snapshots were saved at 30 s intervals.
\FIG{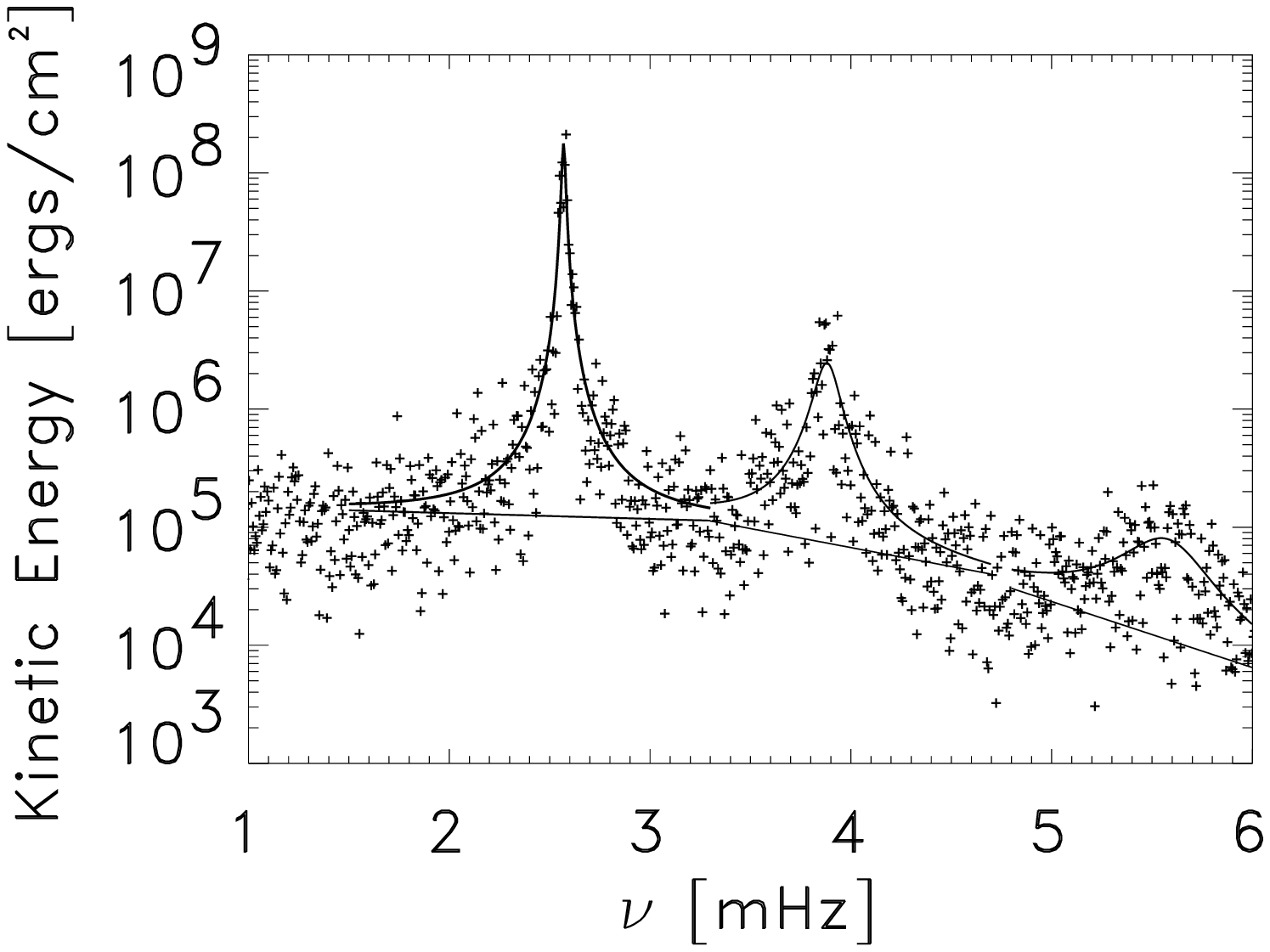}{8.0cm}{
   Kinetic energy, horizontally averaged and integrated over depth.
   Three radial modes are clearly visible.  Least squares, Lorentzian,
   fits to the modes and linear fits to the background noise are 
   superimposed.
}

\begin{center}
\begin{table}
\caption{Mode Properties}
\begin{tabular}{|c|c|c|c|c|c|}  
\hline
$\nu$ & 
$\Delta \nu$ &
$E_{\omega}$ & 
Q & 
$\Gamma$ &
dE/dt \\ 
(mHz) & ($\mu$Hz) & (ergs) & & (s$^{-1}$) & ( erg s$^{-1}$) \\ \hline
2.57 & 19.0 & 3.4 $\! \times \!$ 10$^{26}$ & 135 & 
1.2 $\! \times \!$ 10$^{-4}$ & 4.3 $\! \times \!$ 10$^{22}$ \\ \hline
3.88 & 133 & 2.6 $\! \times \!$ 10$^{25}$ &  29 & 
8.4 $\! \times \!$ 10$^{-4}$ & 2.2 $\! \times \!$ 10$^{22}$ \\ \hline
5.58 & 438 & 2.2 $\! \times \!$ 10$^{24}$ &  13 & 
2.7 $\! \times \!$ 10$^{-3}$ & 6.1 $\! \times \!$ 10$^{21}$ \\ \hline
\end{tabular}
\label{tab_mode_prop} 
\end{table}
\end{center}

The eigenfunctions of the velocity are calculated by taking the
time Fourier transform of the velocity.  To get the real eigenmode,
the transform of the velocity at the frequency of the mode is divided
by its most common phase among all depths.  To reduce the
noise, we average the result over a frequency band, approximately 
equal to the FWHM of the mode, centered on the mode.  The eigenfunctions
are essentially the same as the solar model modes of Christensen-Dalsgaard
using his spherically symmetric model S \citep{JCD+96science}
at the mode frequencies (fig~\ref{figure2}), when the modes are normalized
by the square root of the mode energy, or which is equivalent by their
amplitude at the surface.  Hence, we choose to use the solar model
eigenfunctions, because they are much denser (35 radial modes below the
acoustic cutoff frequency instead of three) and because they are slightly
smoother.
\FIG{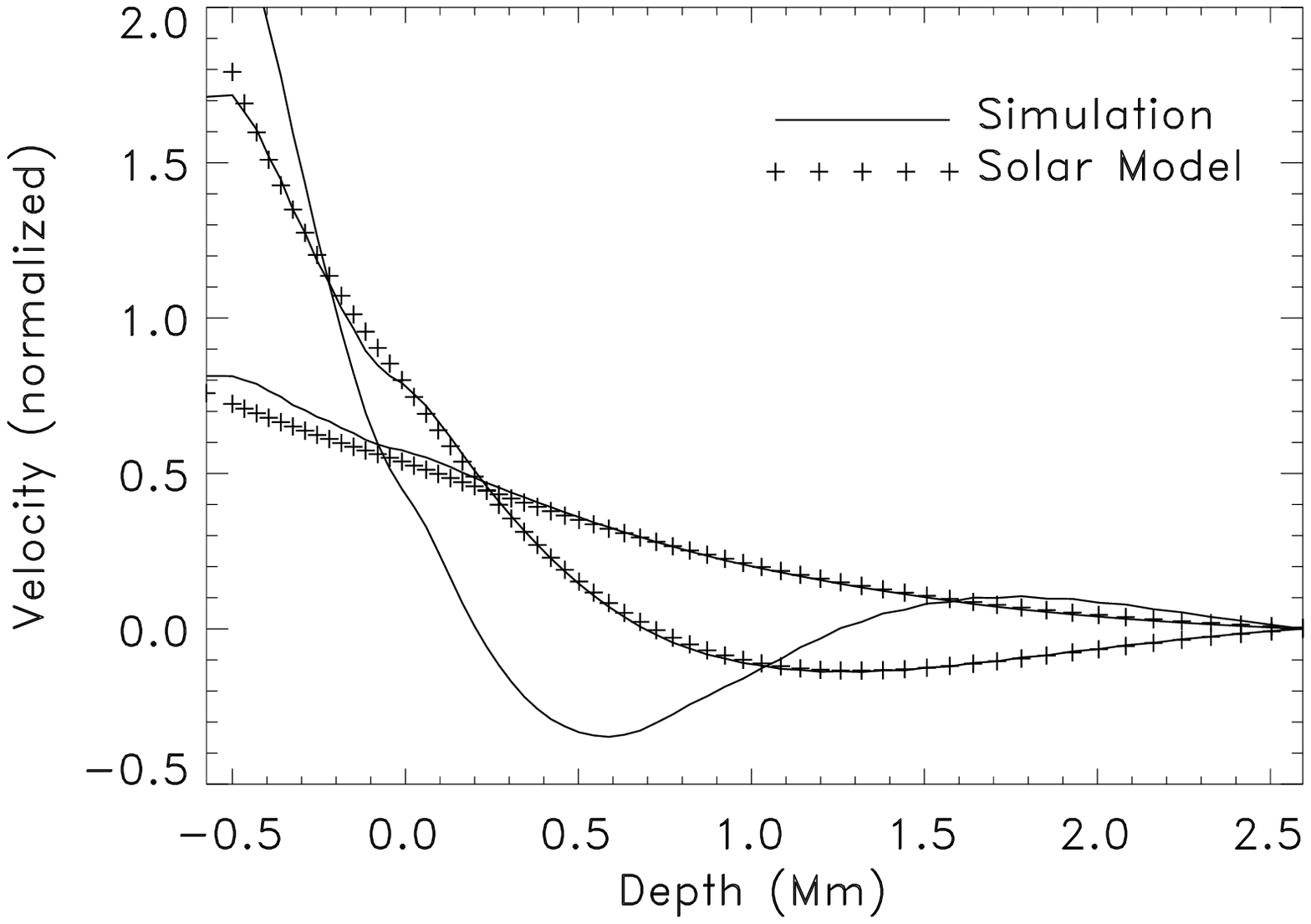}{8.0cm}{
   Velocity eigenmodes of the simulation compared with those of a solar
model \citep[model S of][]{JCD+96science}.  The modes
are normalize by the square root of their energy (eqn~\ref{eq_enorm}).
The solar modes that have nodes at the bottom of the simulation domain 
closely match the simulation modes.  The simulation $p_1$ mode corresponds
to the solar $p_{16}$ mode and the simulation $p_2$ mode corresponds
to the solar $p_{26}$ mode.  The  simulation $p_3$ mode lies above the
acoustic cutoff frequency.
}
Another way of looking at the modes is via their kinetic energy.  Figure
\ref{figure3} is an image of the logarithm of the kinetic energy
as a function of depth and frequency.  The three modes of the computational
domain are clearly seen.  Notice also that there is a continuous pattern
of nodes and antinodes, with the nodes descending in depth as the
frequency increases and new nodes starting at the surface.  This pattern
extends even into the propagating region above the acoustic cutoff
frequency of about 5.3 mHz.
\FIG{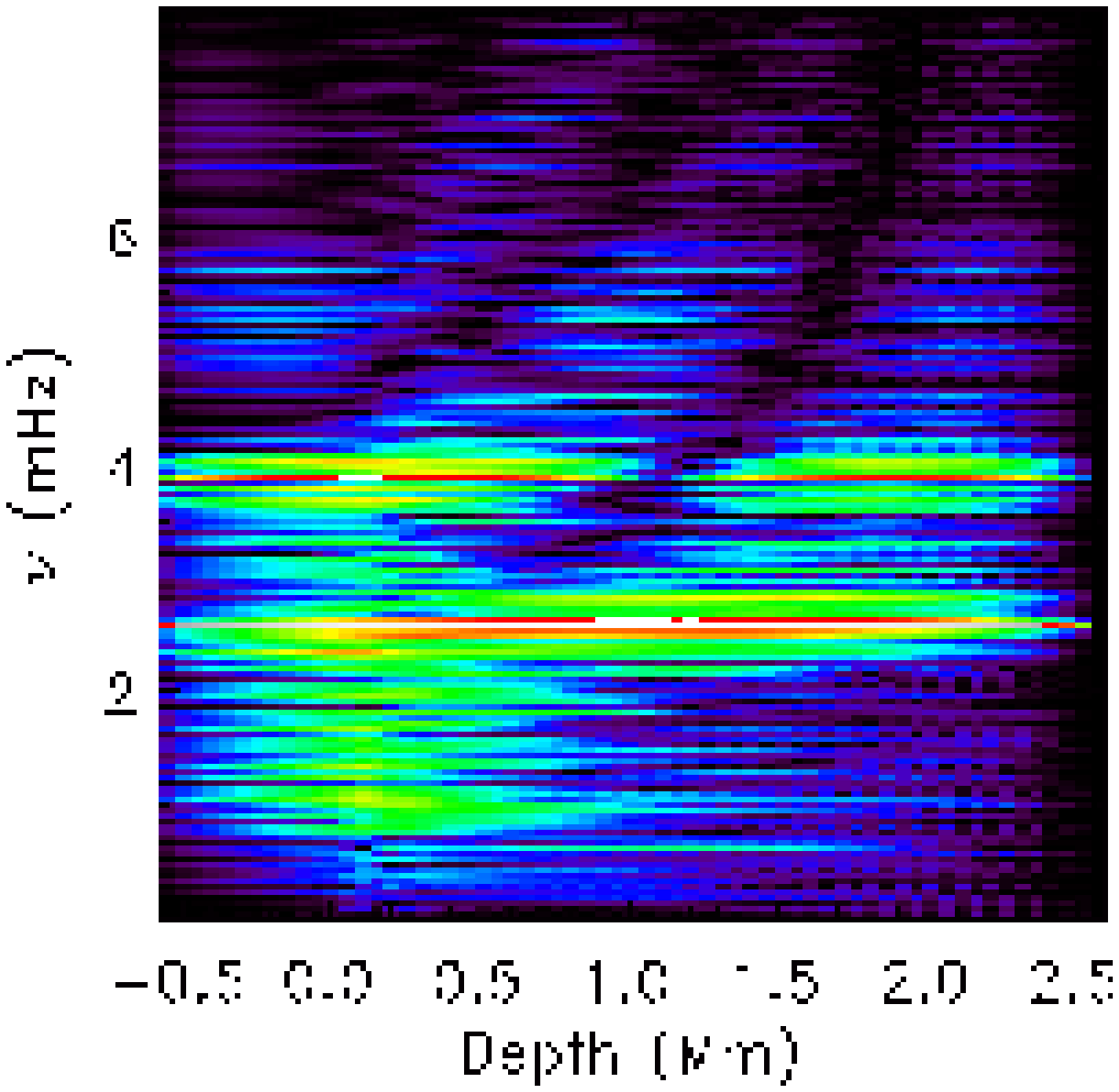}{7.0cm}{
   Logarithm of kinetic energy as a function of frequency and depth.
   The three resonant modes of the simulation stand out as maxima in the
   kinetic energy. A
   regular, continuous pattern of nodes and antinodes in the kinetic energy
   exists both at the resonant modes and between them.
}

The actual mode energy decay rate of the simulation modes, which on
average is 
equal to their excitation rate, is obtained by multiplying the energy of
each mode by its decay rate $\Gamma = 2 \pi \Delta \nu_{\rm FWHM}$, which is
its radian line width and is obtained from the fit to the modes
(fig~\ref{figure1}).

The total mode energy is the
sum of the mode energies minus a fit to the background over the
frequency range where the modes are significantly above the background
multiplied by the area (36 Mm$^2$) of the simulation domain
(fig~\ref{figure1}).  

Eqn~(\ref{eq_edot}) is used to predict the mode 
excitation rate.  The non-adiabatic total (gas + turbulent) pressure
fluctuations are calculated directly from the simulation by first
averaging the gas pressure, turbulent pressure and density over
horizontal planes at each saved snapshot.  These are then interpolated
to the Lagrangian frame at each time.  The non-adiabatic pressure fluctuations
are calculated as in eqns~\ref{eqn_pnad} and \ref{eqn_pnad_fluct}.
The oscillation modes
of the domain are essentially adiabatic and do not affect the
non-adiabatic pressure fluctuations as can be seen from its featureless,
noisy spectrum (Fig~\ref{figure4}).  Thus the non-adiabatic pressure
work is due primarily to the convection.

\FIG{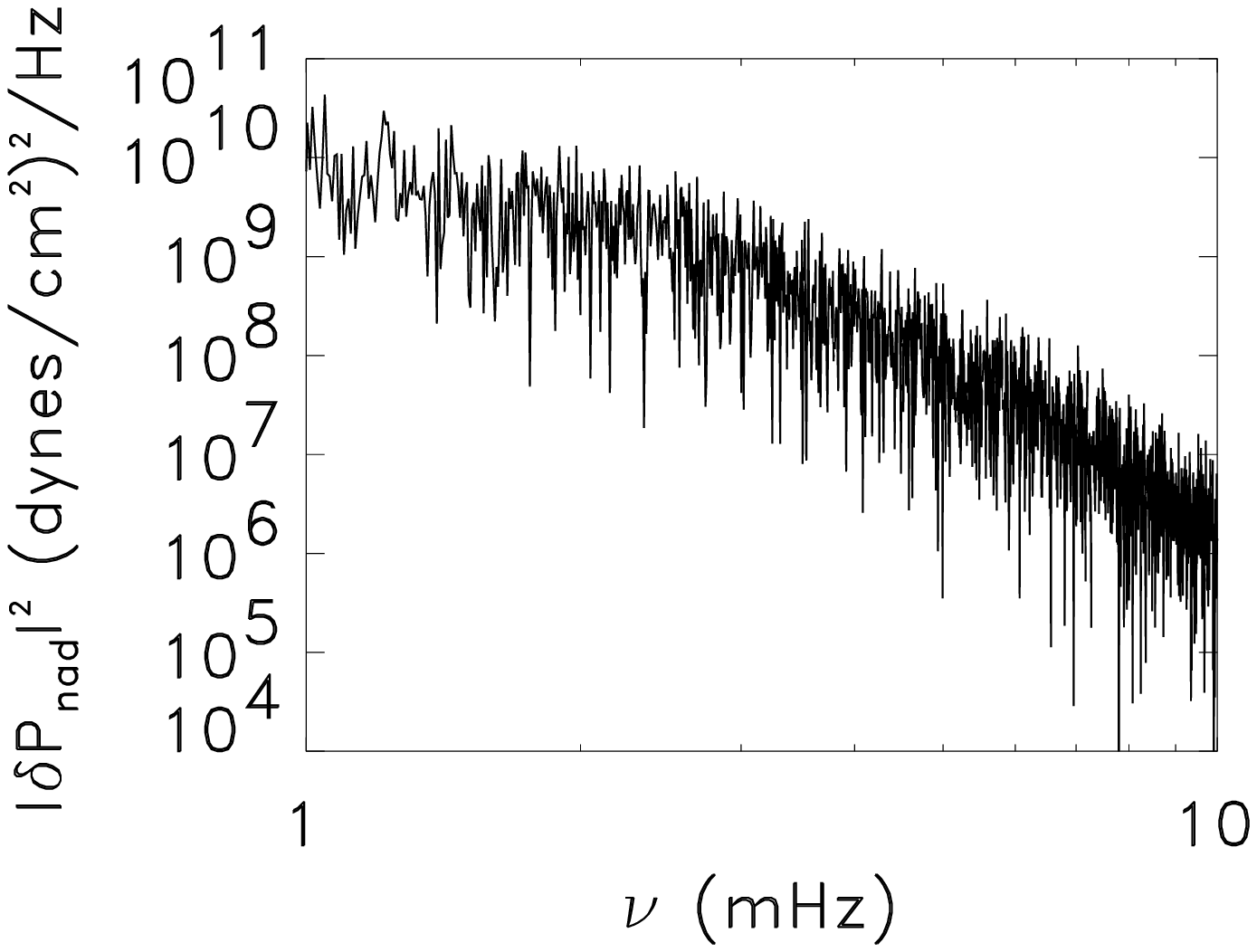}{8.0cm}{
Non-adiabatic pressure spectrum at the a depth of 100 km.  Note that 
it is featureless even
at the frequencies of the simulation modes.  Hence, the non-adiabatic
pressure is primarily due to random convective processes.
}
\FIG{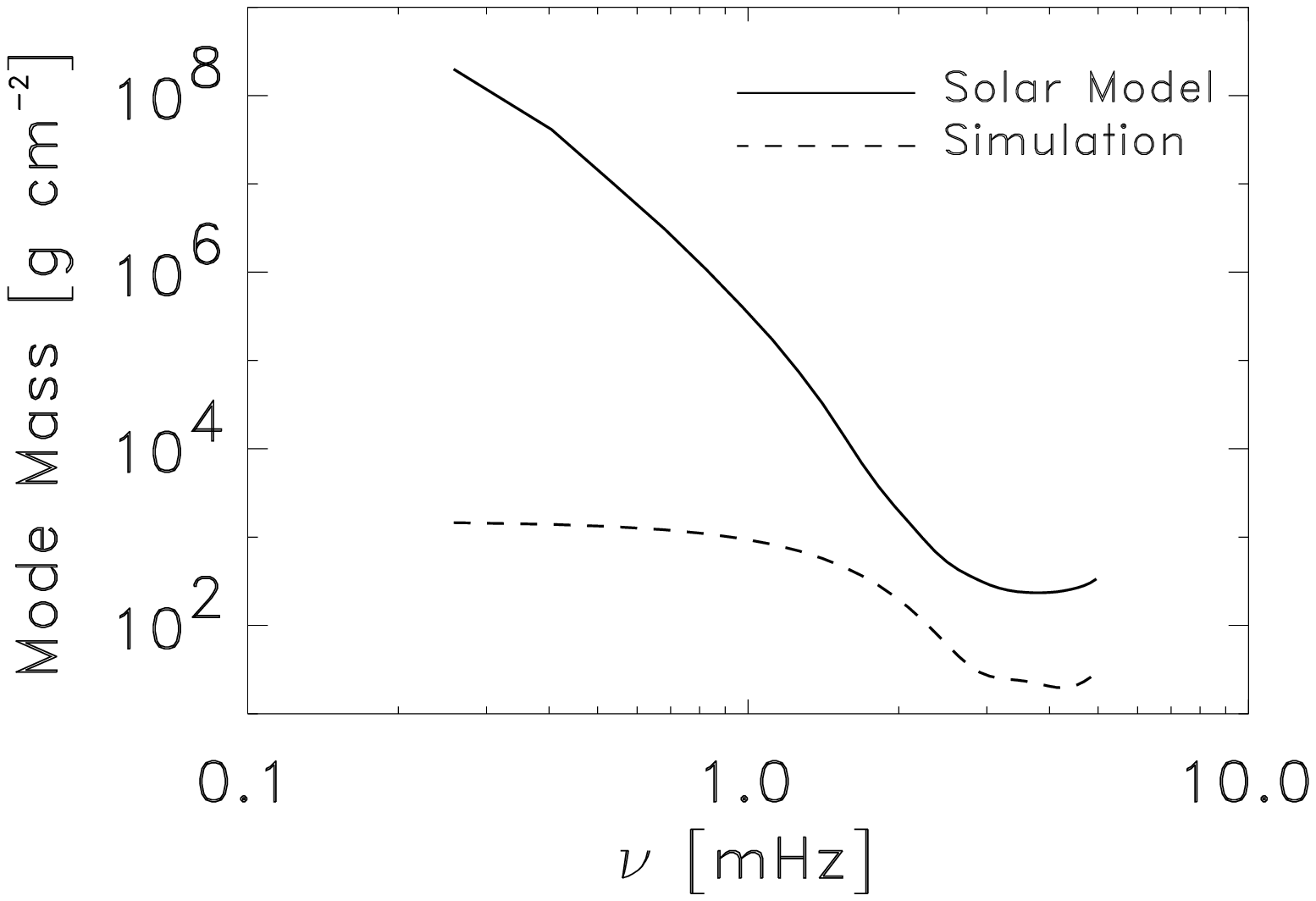}{8.0cm}{
   Mode mass in the simulation domain and the Sun.  Mode mass decreases
   with increasing frequency because higher frequency modes are more
   concentrated toward the surface than low frequency modes.  Because
   the solar modes extend below the bottom of the simulation domain,
   they have a larger mode mass.  Because the simulation domain is
   shallow, the mode mass becomes nearly constant at low frequency
   where the eigenfunctions become nearly constant within the depth of
   the simulation.
}
\FIG{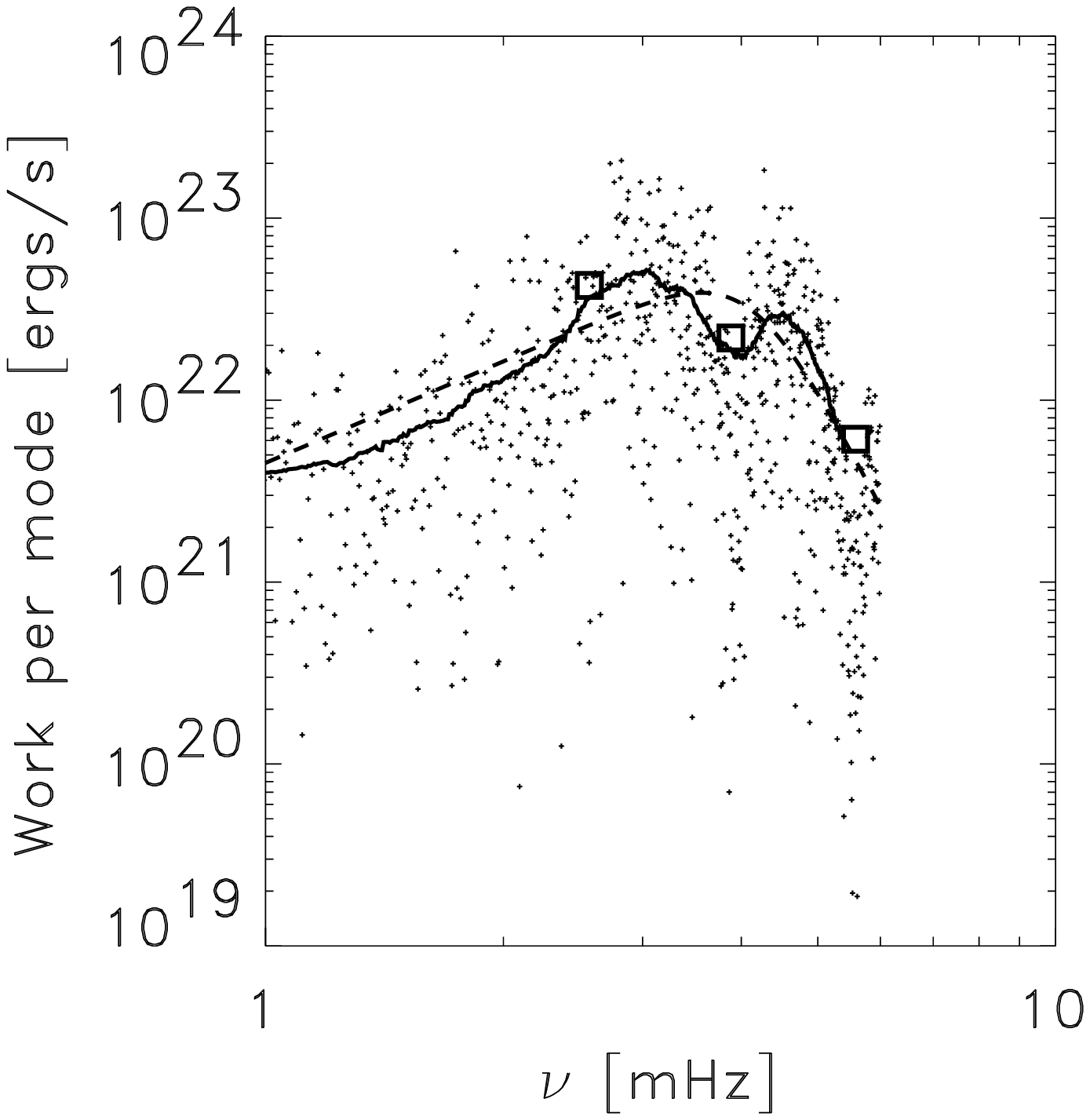}{8.0cm}{
   Rate of stochastic energy input to the simulation modes (squares)
   compared to the predicted excitation rate (small crosses) from the
   work of non-adiabatic pressure fluctuations on the modes
   (eqn~\ref{eq_edot}).  The solid line is a running boxcar average and
   the dashed line a two power law fit over the entire frequency
   range.  The excitation rate is larger than the solar case
   because the mass of the modes in the simulation is less than the
   mass of the modes in the Sun.  The near constancy of simulation mode
   mass at low frequency leads to a much slower decrease of excitation
   with decreasing frequency than occurs in the Sun.  
}

The mode compression is calculated
from the Christensen-Dalsgaard modes (since these are essentially 
identical with the simulation modes at the resonant frequencies) normalized
by the square root of their energy in the simulation domain.  Because
the simulation domain is shallow, while, especially at low frequency, the
solar modes have substantial amplitude below the computational domain,
the mode mass in the computational box is significantly smaller than the
actual mode mass of the solar modes (Fig~\ref{figure5}).  

\FIG{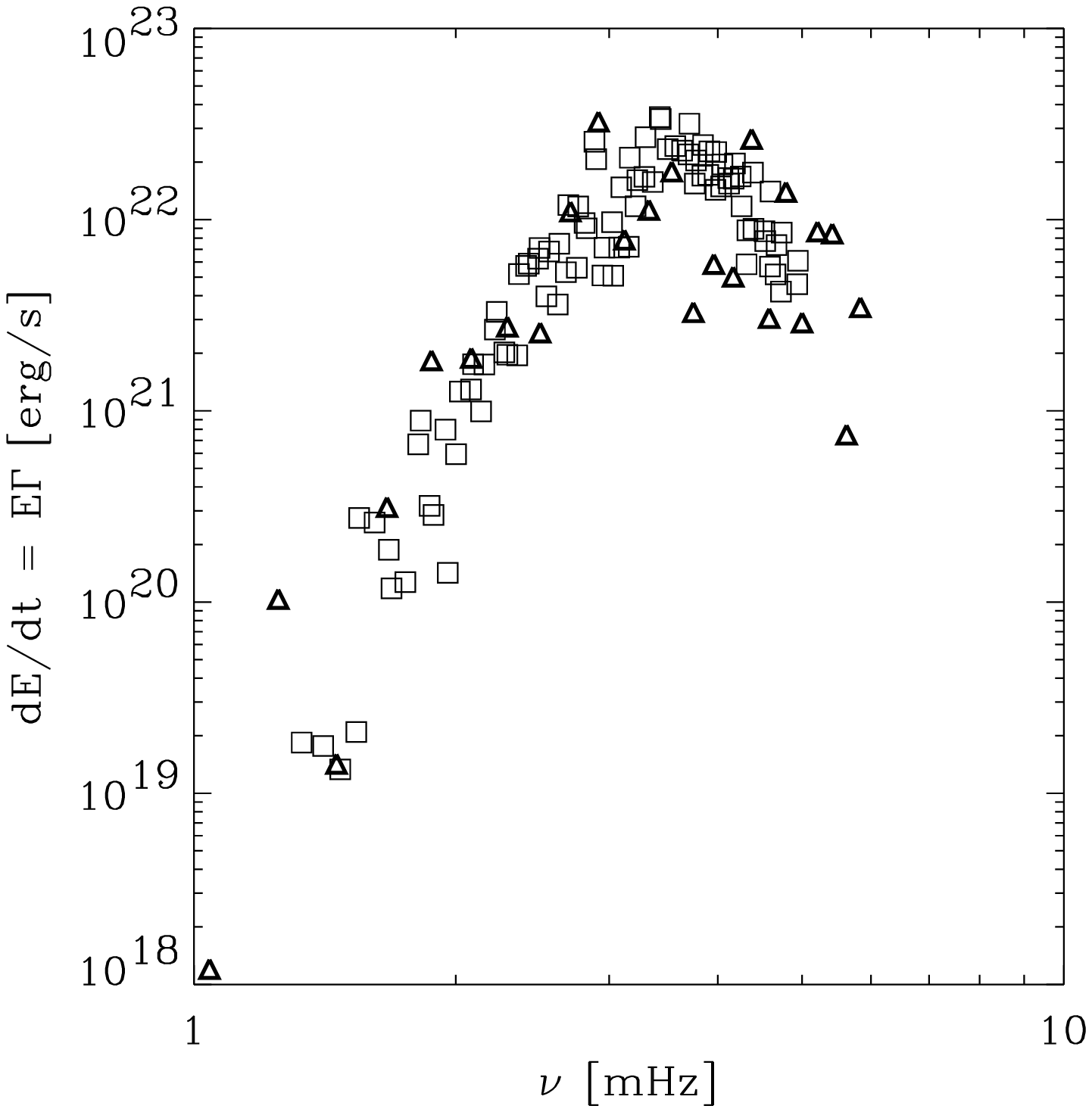}{8.0cm}{
   Rate of stochastic energy input to modes for the entire solar surface  
   (simulation=triangles, observations=squares) from Roca Cortes (1999), 
   based on observed mode velocity amplitudes and line widths from GOLF,
   for modes with $\ell = 0 - 3$, which are all nearly radial close to the
   surface.  The rate of energy input to the solar modes is smaller
   than to the simulation modes by the ratio of the mode mass of the solar
   modes to the mode mass in the simulation domain 
   (fig~\ref{figure5}).
}

Finally, we compare the actual mode energy decay rate with the
predicted rate of work by convection on the modes given by eqn
(\ref{eq_edot}) in figure~\ref{figure6}.  The squares are the actual
mode energy decay rates for the three resonant modes of the box.  The
solid line is the running mean of the predicted work and the dashed
line is a smooth two power law fit to the predicted work.  The
agreement is very good, but not perfect.  There are significant
deviations from the power law fit in the neighborhood of the modes and
these are reflected in the actual mode decay rate.  This is a phenomena
that still remains to be explained.  Notice also that the decrease in
work toward low frequencies is much slower than for the Sun
(Fig~\ref{figure7}).  The reason is the near constancy of the
mode mass within the simulation domain at these low frequencies
compared with the steeply rising mode mass with decreasing frequency on
the Sun.  This application of our excitation rate formula 
to the modes that are excited in our simulation verifies that
the formula is correct and can be applied to the Sun.

\section{Mode Excitation: Sun}

\subsection{Excitation Spectrum}
\FIG{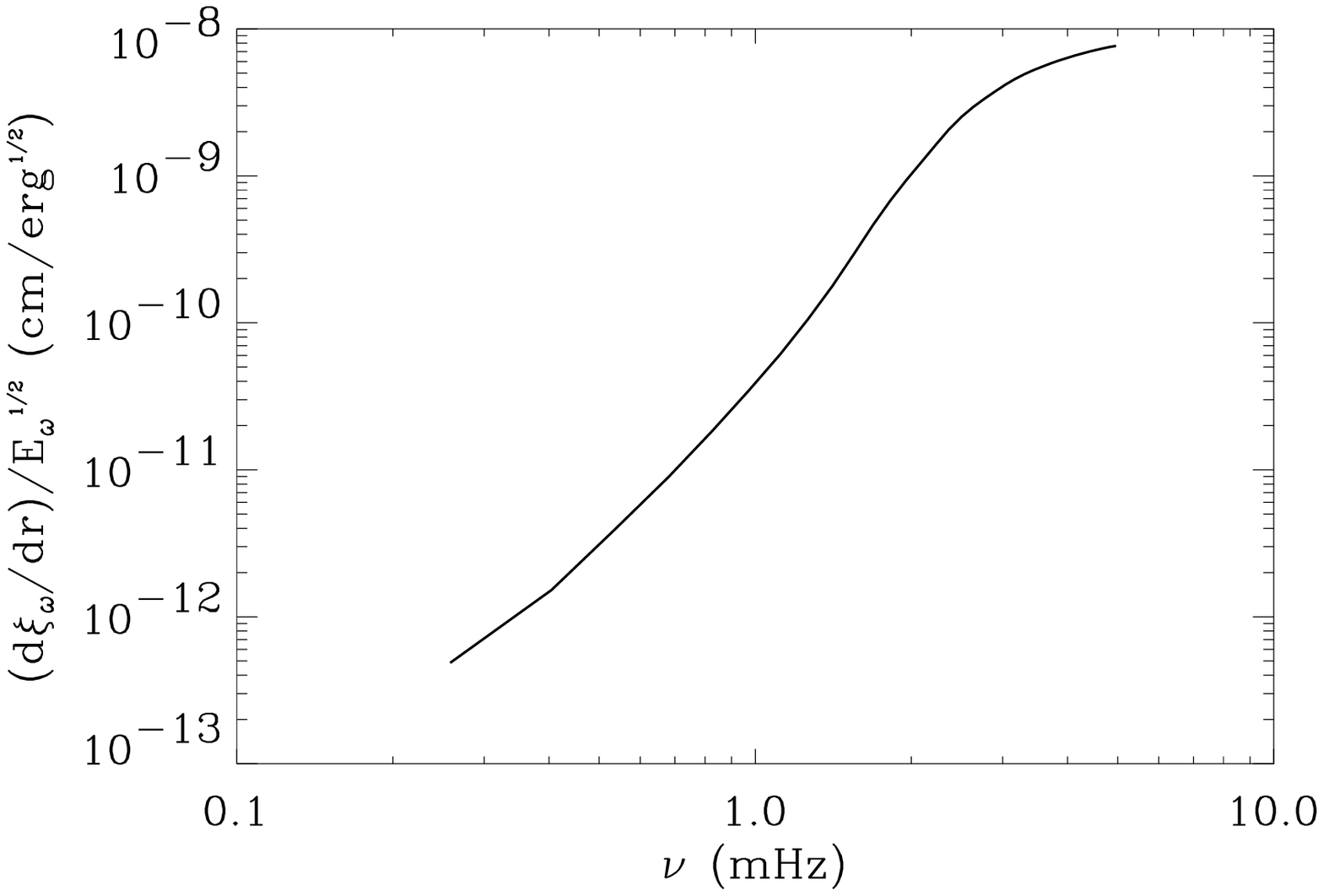}{8.0cm}{
   The mode factor in the work integral:
   ${{\partial \xi_{\omega}} \over {\partial z}} / E_{\omega}^{1/2}$.
   Excitation decreases at low frequency due to mode behavior, in part
   because the radial wave vector is approximately $k=\omega^2/g$ and in
   part because the mode mass increases with decreasing frequency
   (Fig~\ref{figure5}). 
}

The excitation rate for solar modes as a function of frequency is shown in
Fig.~\ref{figure7}.  To obtain these results we used the shorter but
higher resolution $125^2 \times 82$ simulation because it has more high
frequency power.  The magnitude and frequency dependence we find
for our 6 Mm square box is very close to the observed values for the
entire sun.  This means that the pressure fluctuations must be
uncorrelated on larger scales, so there is no extra driving
contribution.  
\nocite{RocaCortes+99}

What produces this frequency dependence?  The low frequency behavior is
controlled by the nature of the eigenmodes and the high frequency
behavior is controlled by the nature of convection.
The work integral (Eqn. \ref{eq_edot}) contains a factor pertaining 
to the modes: the radial gradient of the mode displacement normalized 
by the square root of the
mode energy (which makes it independent of the mode amplitude). This is small
at low frequencies and increases with frequency approximately as $\nu^{3.5}$
(Fig~\ref{figure8}).  Part of this dependence is due to the radial gradient of
the displacement.  The mode dispersion relation is $\omega = \surd(gk)$, so
$k=\omega^2/g$, which accounts for two powers of the frequency.  The
remainder of the frequency dependence is due to the mode mass, which
decreases with increasing frequency because higher frequency modes 
are more concentrated toward the surface (Fig.~\ref{figure5}).

The mode excitation decreases at high frequency because the pressure
fluctuation power decreases with increasing frequency roughly as $\nu^{-4}$
(Fig.~\ref{figure9}).  Convective motions, whose power
decreases at small scales and high frequencies, produce the gas and
turbulent pressure fluctuations.  This then leads to a similar high
frequency decline in the pressure spectrum.

Total pressure fluctuations are small at low frequency
because the atmosphere is nearly in hydrostatic equilibrium.  However,
the turbulent pressure fluctuations are largest at low frequency because
they are a convective effect and convection is a longer time scale
phenomena.  Hence, the gas pressure fluctuations must also be large and
out of phase with the turbulent pressure fluctuations at low frequency
in order to produce small total pressure fluctuations  
(Fig.~\ref{figure10}).  The turbulent
pressure is small compared to the gas pressure ($\sim$ 15\%), but the
magnitude of fluctuations are comparable for the gas and turbulent
pressures, so they can indeed cancel each other.

\FIG{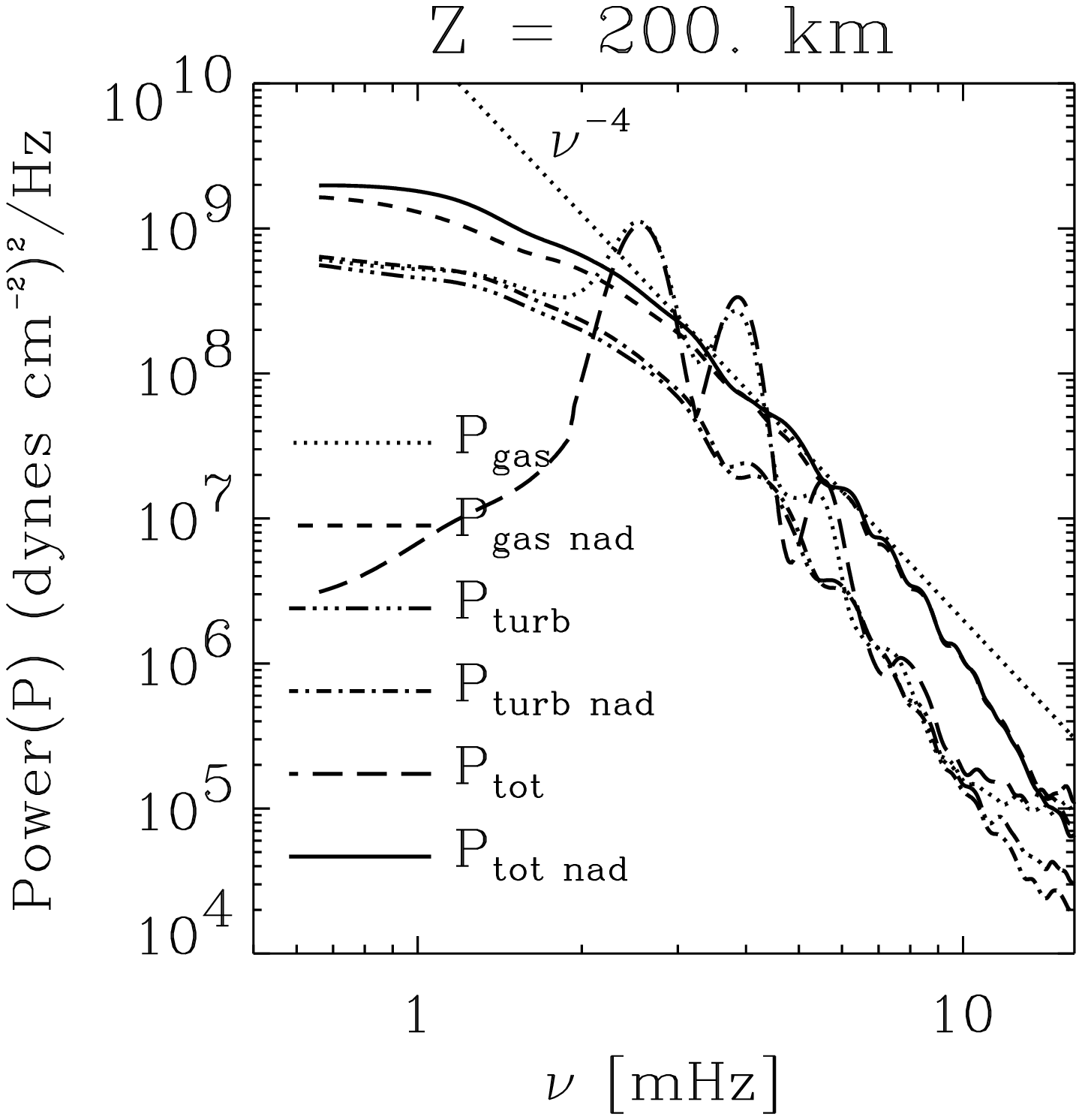}{8.0cm}{
   Spectrum of pressure fluctuations at a depth of 200 km, smoothed
   with running boxcar mean.  The non-adiabatic gas pressure
   fluctuations exceed the non-adiabatic turbulent pressure
   fluctuations by about a factor of four, but they become comparable
   in the peak driving range at larger depths.  The local maxima in
   the total pressure fluctuations at 2.6, 3.9 and 5.5 mHz are due to
   the resonant modes in the simulation.  The non-adiabatic pressure
   varies smoothly across these resonant frequencies, indicating it is
   primarily due to convection.  The pressure fluctuation power 
   decreases roughly as $\nu^{-4}$ at high frequency, because it is due
   due to stochastic convective motions which decrease in power at 
   high frequency.
}
\FIG{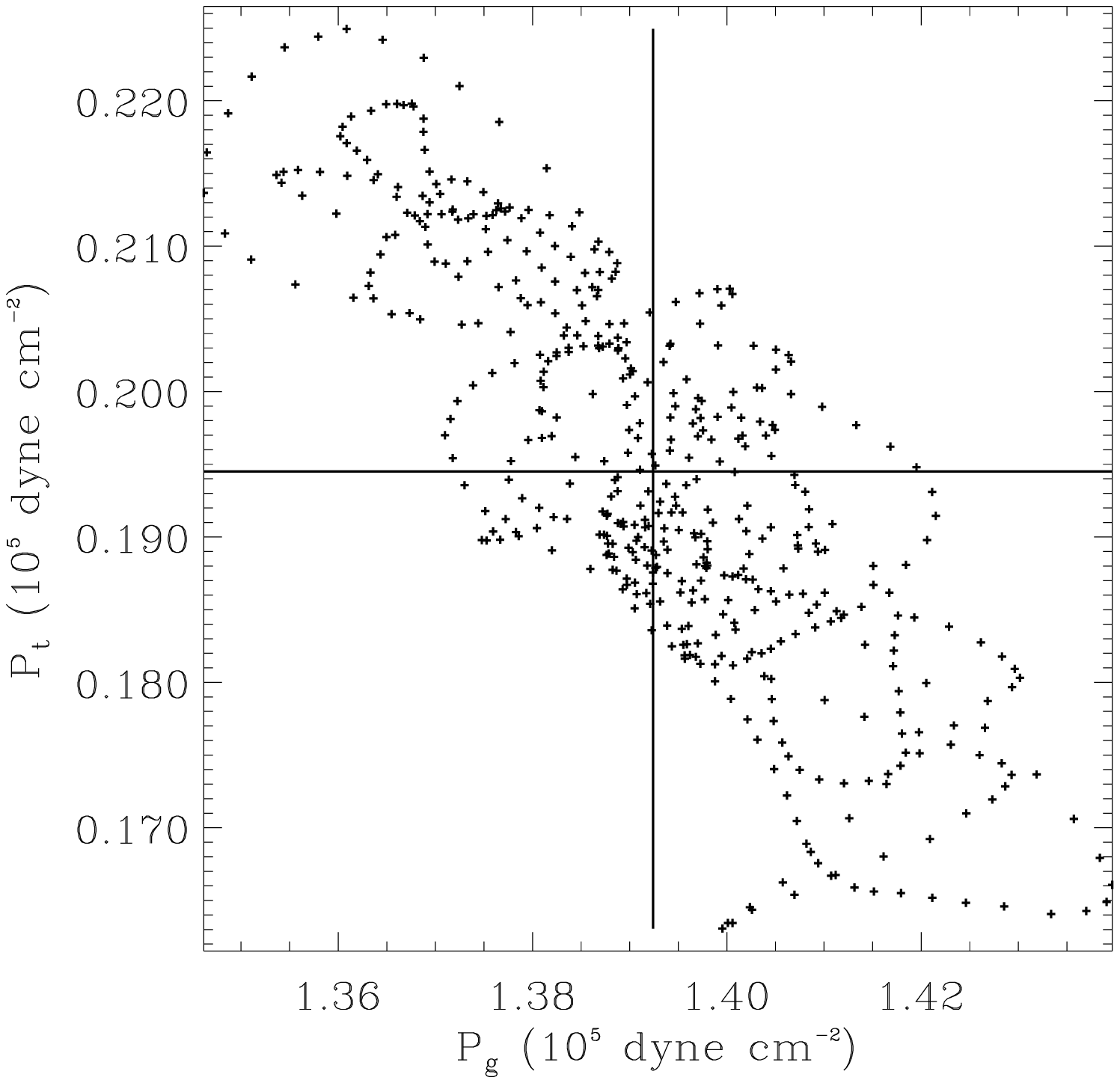}{8.0cm}{
   Correlation of turbulent and gas pressure at the surface.  The 
   turbulent pressure is
   only about 1/6 the magnitude of the gas pressure near the surface but
   the magnitude of their fluctuations are similar. 
}
\FIG{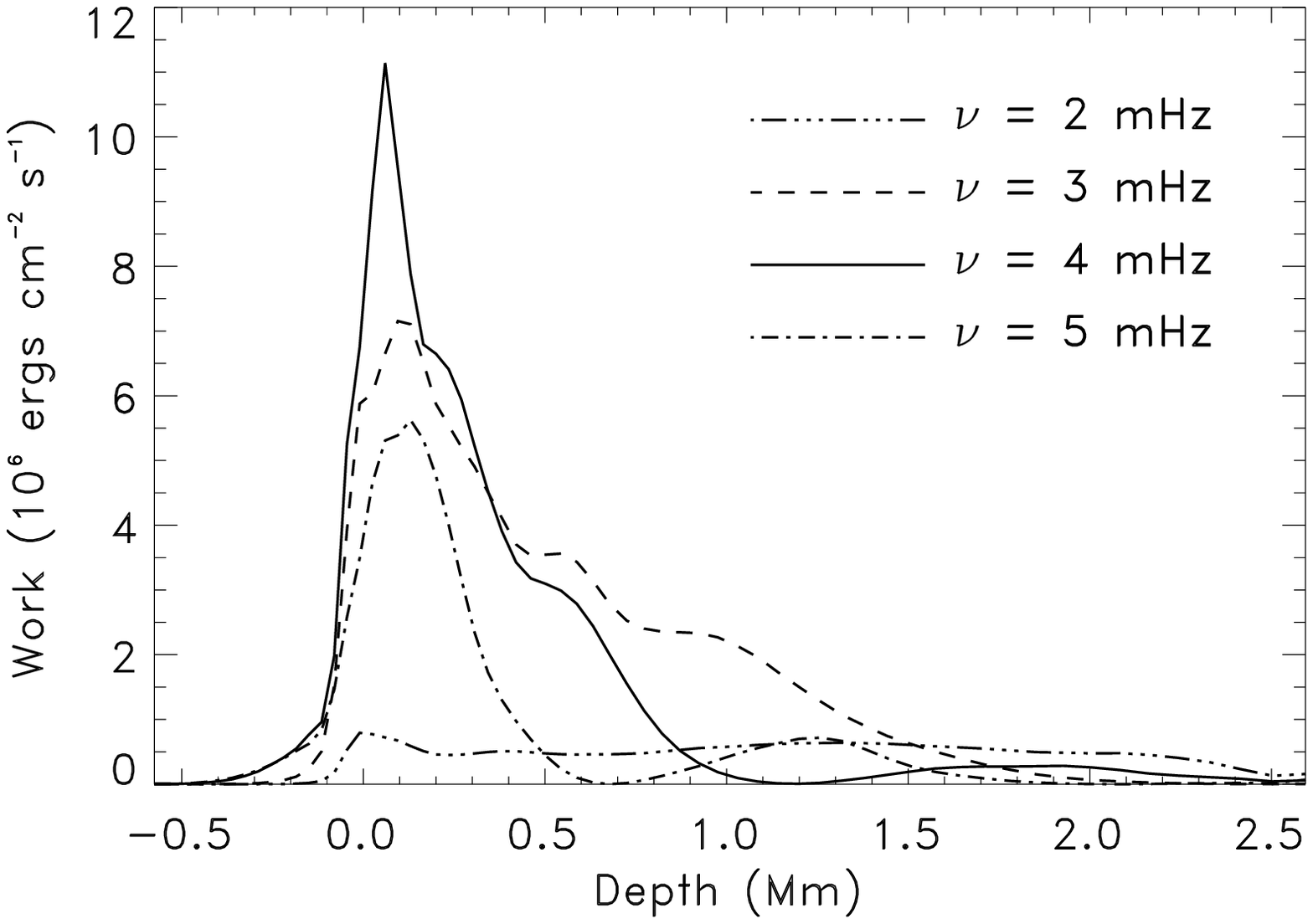}{8.0cm}{
   Integrand of the work integral,
   ${\omega^2 \left | \delta P^*_{\omega} {{\partial \xi_{\omega}}
   \over {\partial z}} \right |^2} / {8 \Delta \nu E_{\omega}}$,
   (eqn ~\ref{eq_edot}) at 2, 3, 4, and 5 mHz as a function of depth.  
   At frequencies where the driving is large, the integrand is 
   significant only within 500 km of the surface. 
}

\subsection{Excitation Location}

At what depth does the driving occur?  Consider the integrand of the
work integral at different frequencies (Figs.~\ref{figure11} 
and \ref{figure12}).  At
low frequencies the integrand amplitude is similar over an extended 
depth range, but it is small.  Where the integrand is large, in the
region of peak driving around 3-4 mHz, the integrand becomes
concentrated very close to the surface and most driving occurs between
the surface and 500 km depth.  At still higher frequencies the
integrand becomes small again and even more concentrated near the
surface.
\FIG{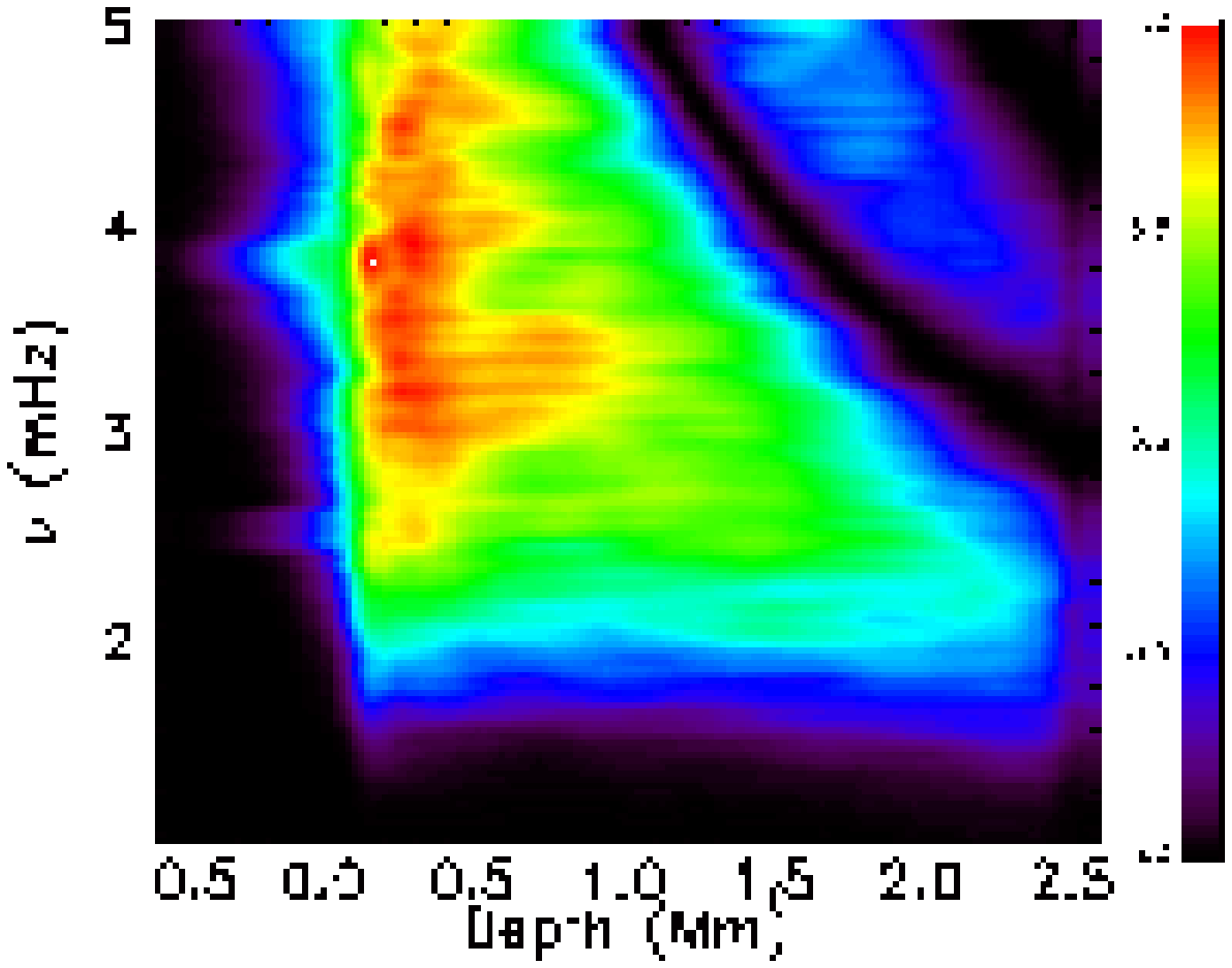}{8.0cm}{
   Logarithm (base 10) of the work integrand,
   ${\omega^2 \left | \delta P^*_{\omega} {{\partial \xi_{\omega}}
   \over {\partial z}} \right |^2} / {8 \Delta \nu E_{\omega}}$,
   (eqn ~\ref{eq_edot}) (in units of ergs/cm$^2$/s) 
   as a function of depth and freqency.  The work
   is concentrated close to the surface in the peak excitation range
   (3 - 4 mHz) and at higher frequencies.
}
\FIG{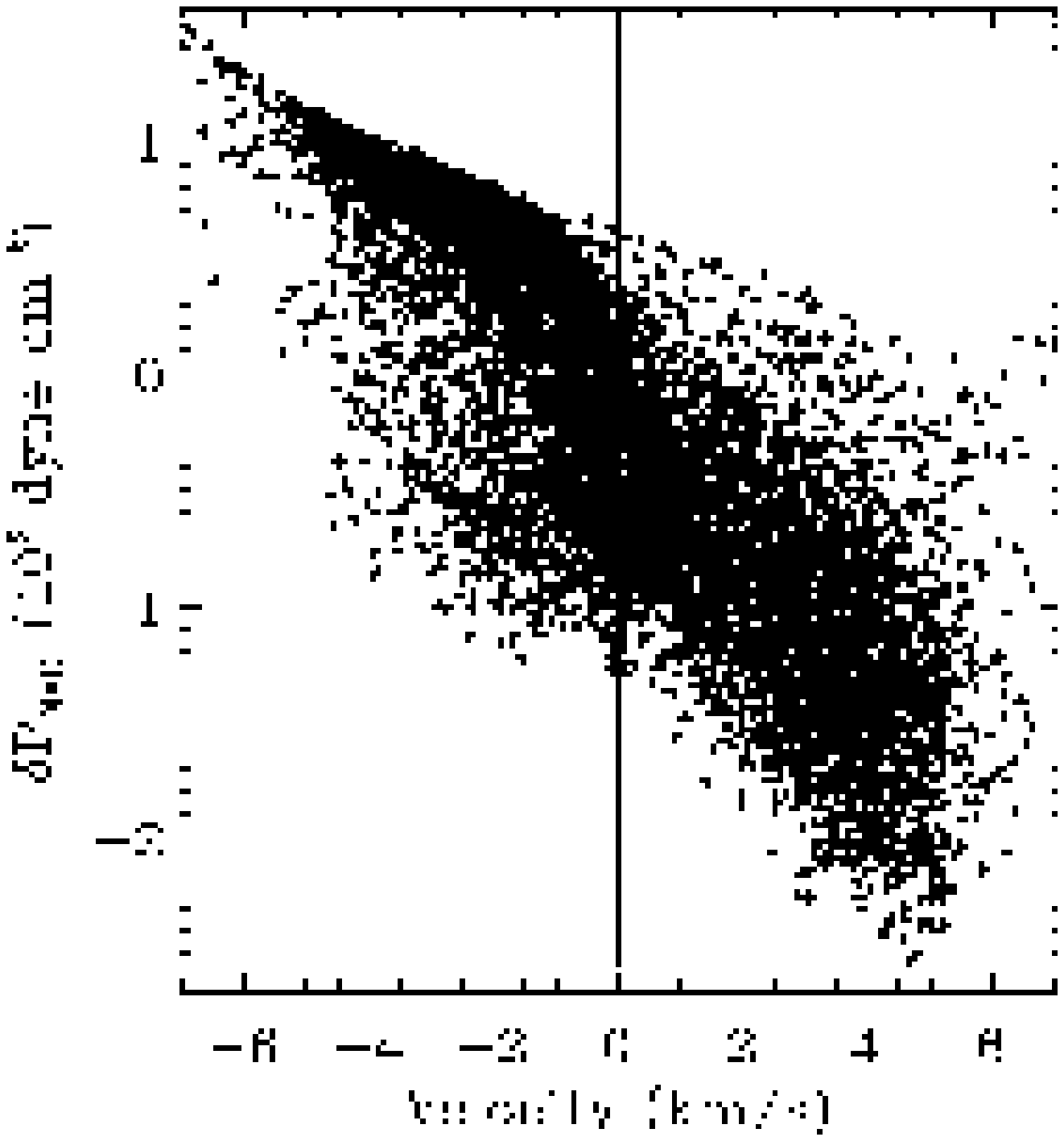}{7.0cm}{
   Correlation of non-adiabatic pressure with vertical velocity at the surface.
   The large negative fluctuations in the non-adiabatic pressure occur where
   the velocity is positive (downward) in the intergranular lanes.
}
\FIG{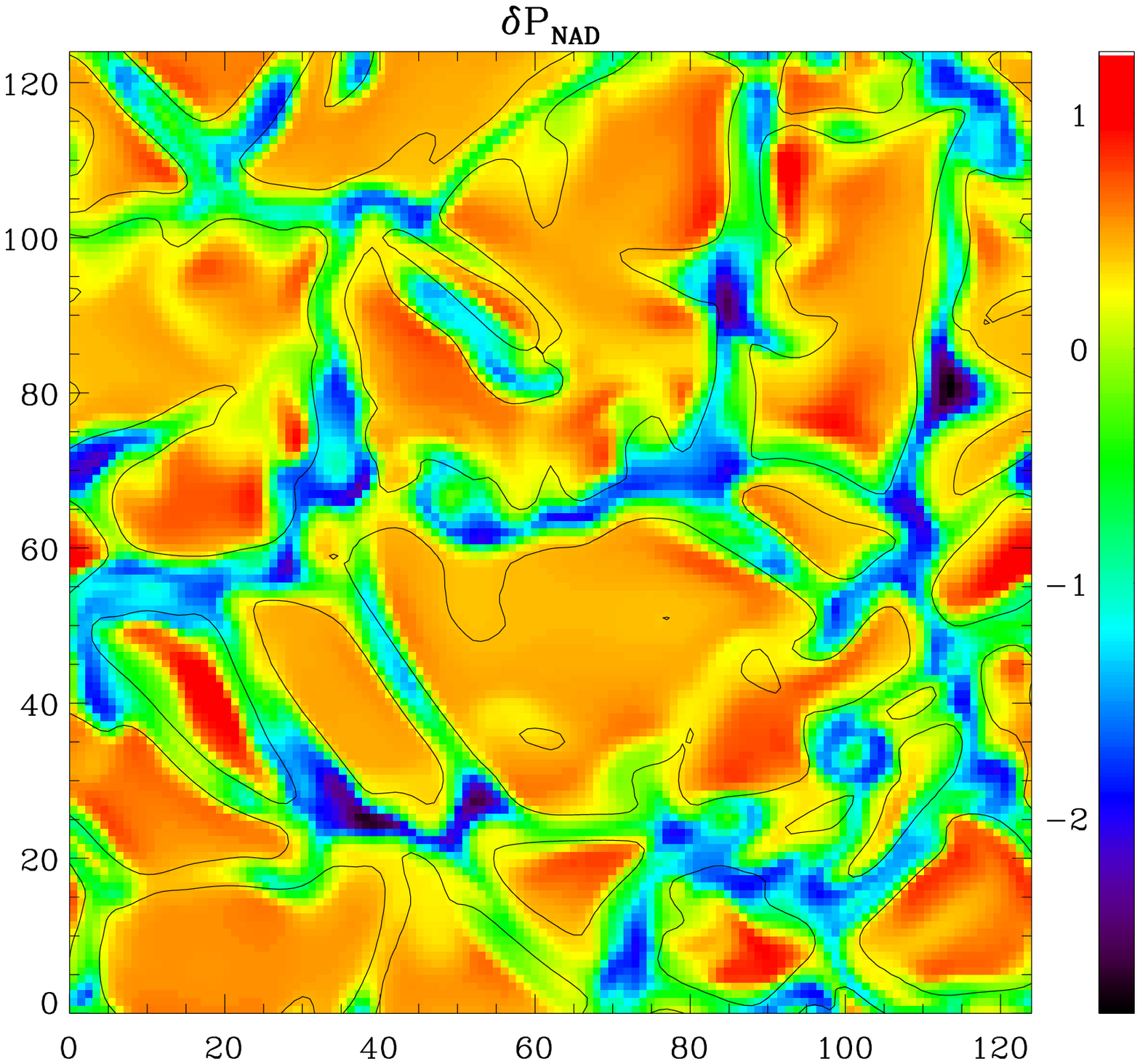}{6.5cm}{
   Image of the non-adiabatic total pressure fluctuations at depth 100
   km, with the contours of zero surface velocity to outline the granules.  
   The units of the pressure are $10^3$ dynes/cm$^2$.
   Large, negative pressure fluctuations occur in the intergranular lanes. 
   Positive pressure fluctuations are half as large and occur inside the 
   granules.
}
\FIG{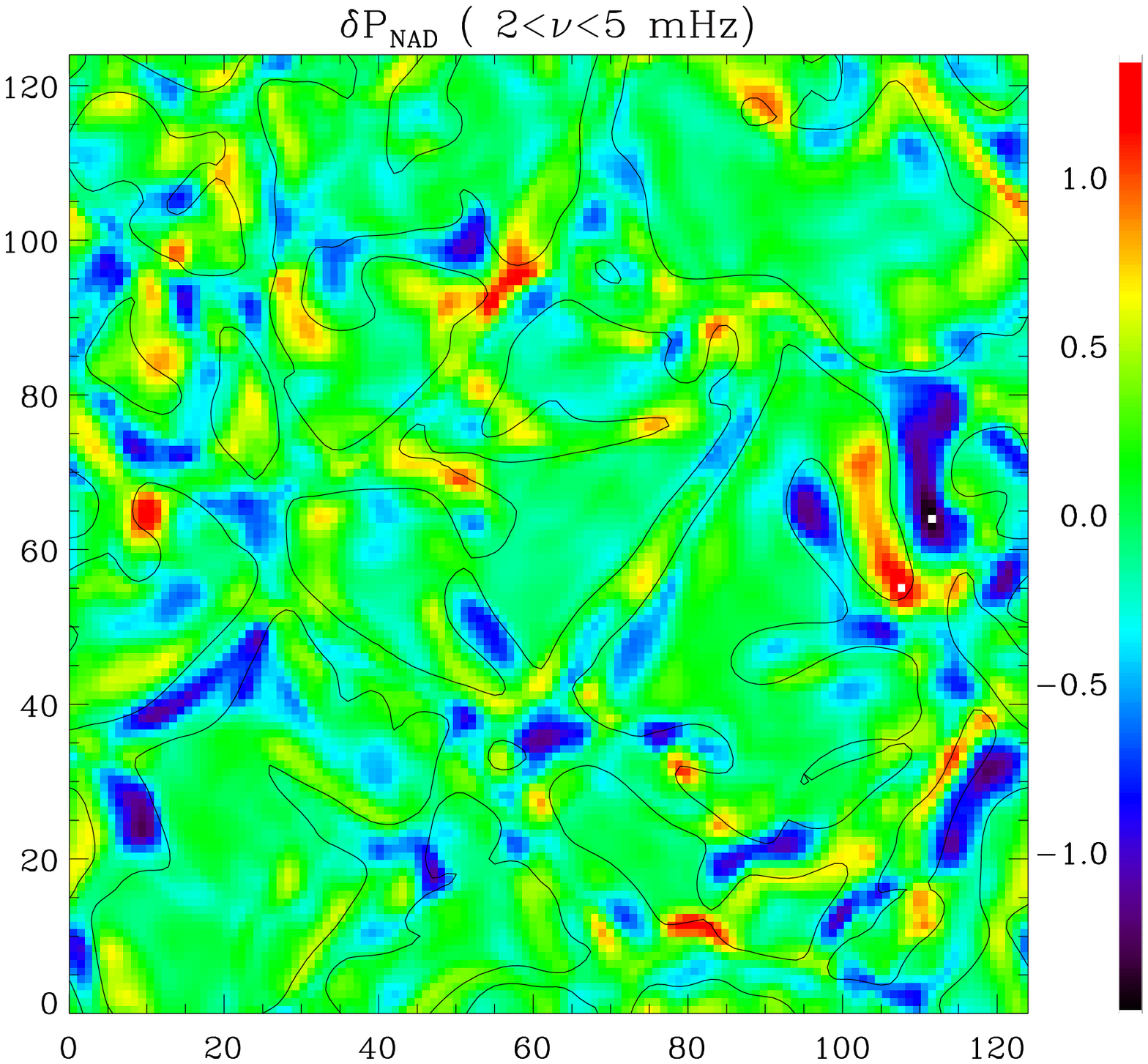}{6.5cm}{
   Image of non-adiabatic pressure fluctuations at 100 km depth in the
   2 -- 5 mHz range, with the contours of zero velocity at the surface
   to outline the granules.  The units of the pressure fluctuations are
   10$^3$ dyne cm$^{-2}$.  In this frequency range, where the driving
   is maximal, the largest $\delta P_{\rm nad}$  occur at the edges of
   granules and inside the intergranular lanes.
}
\FIG{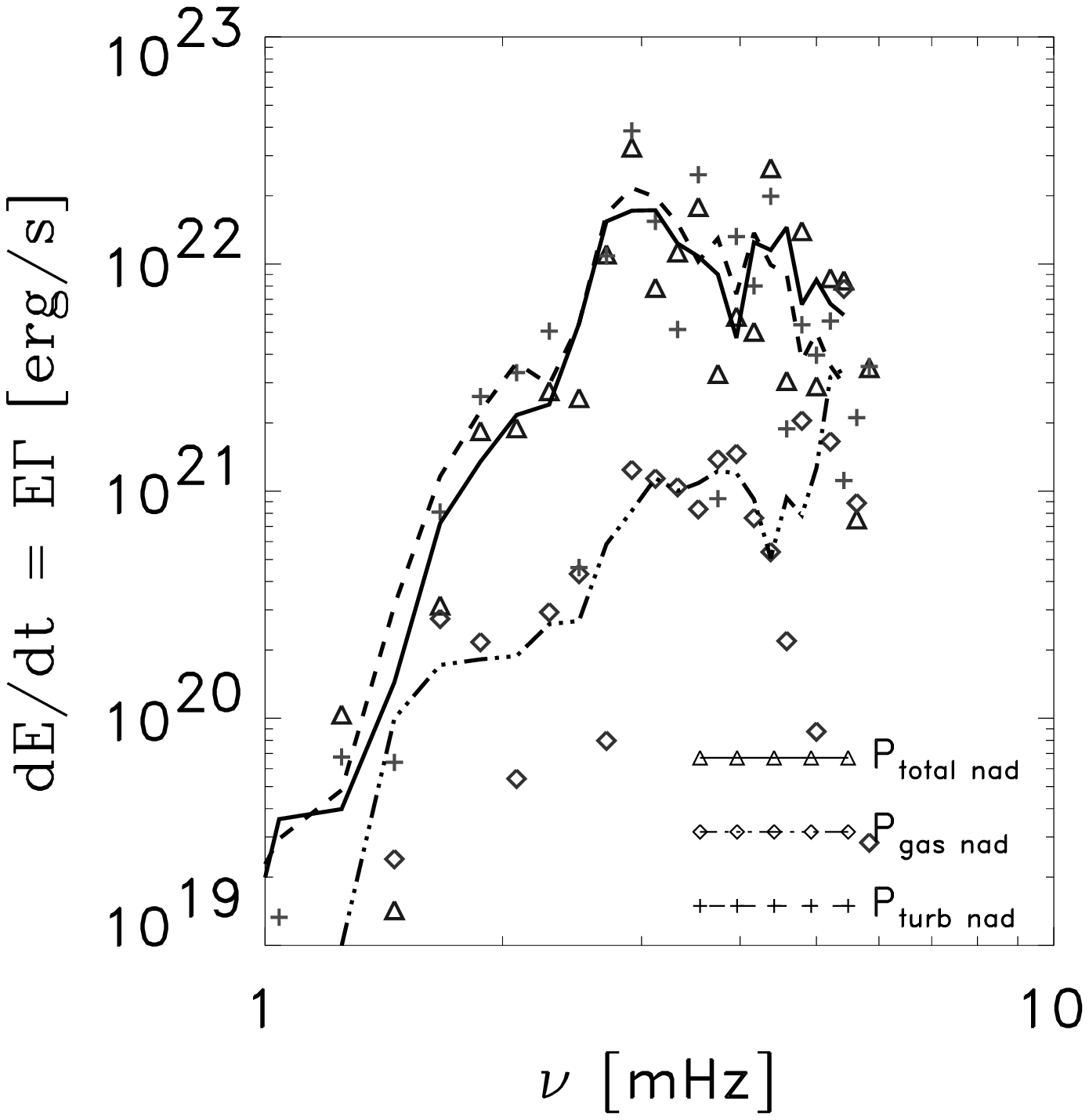}{8.0cm}{
   Rate of stochastic energy input to modes for the entire solar surface,
   showing the individual contributions of the non-adiabatic gas and 
   turbulent pressure to the work of the total non-adiabatic pressure.
   Most of the driving in the peak driving range comes from the
   turbulent pressure.
}
\FIG{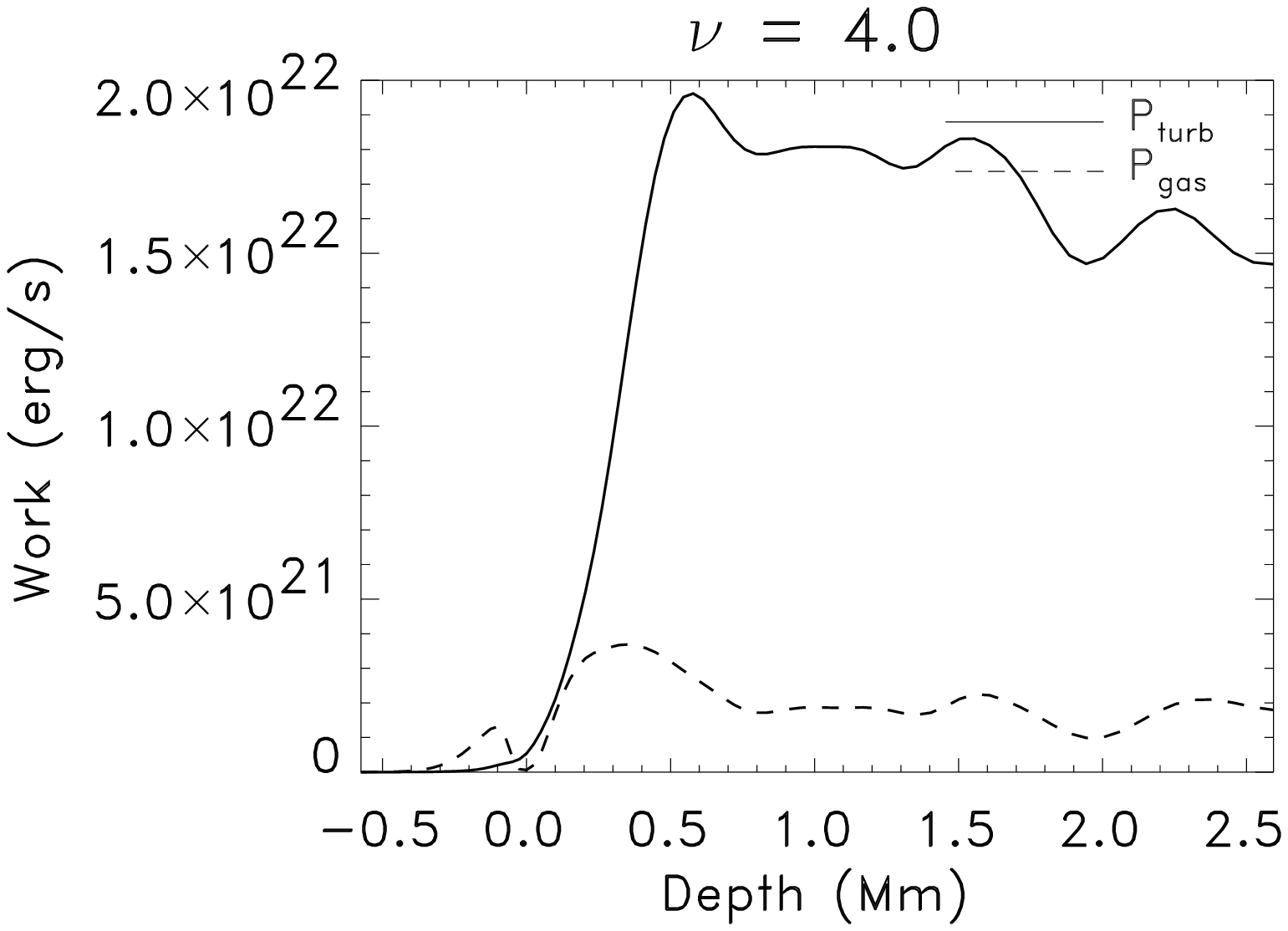}{8.0cm}{
Mode driving at 4 mHz evaluated from the surface to depth z, showing
the individual contributions of the non-adiabatic gas and turbulent 
pressure.  Close to the surface the contributions of the two are
comparable, but there is little contribution from the gas pressure
below 200 km depth, while the turbulent pressure work is significant
down to 500 km depth.
}
\FIG{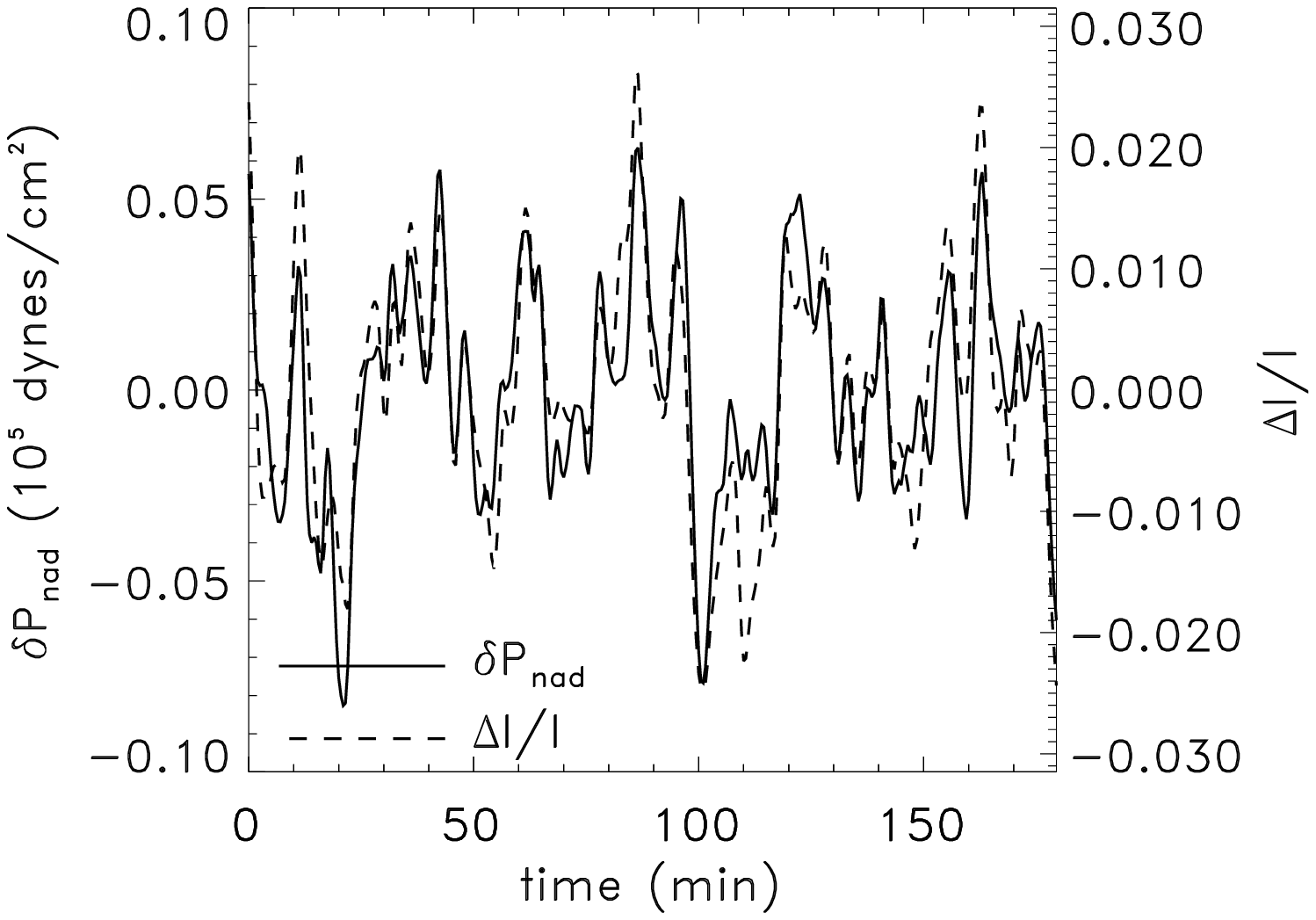}{8.0cm}{
   Horizontally averaged non-adiabatic pressure at the surface and 
   emergent intensity variation in time.  
   They are tightly correlated indicating that radiative cooling at the 
   surface is the source of the non-adiabatic pressure fluctuations
   there..
}
\FIG{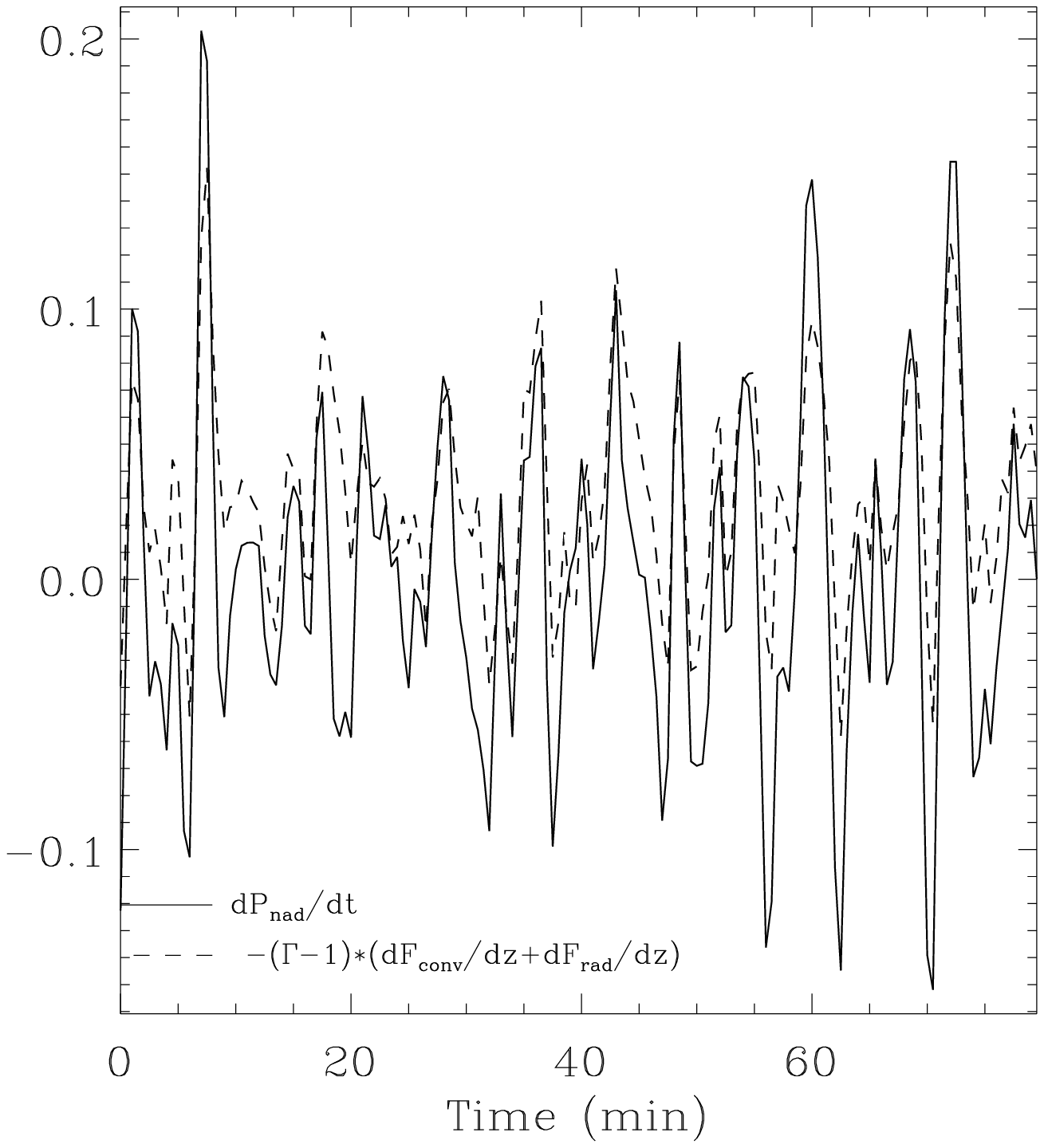}{8.0cm}{
   Divergence of the convective and radiative fluxes, at z=100 km,
   multiplied by ($ \Gamma_3-1$) compared to the time derivative of the
   non-adiabatic pressure at the surface.  The units are 10$^3$
   ergs/cm$^3$/s.  The rate of change of non-adiabatic pressure closely
   follows the divergence of the net flux but has a slightly larger
   amplitude.
}
\FIG{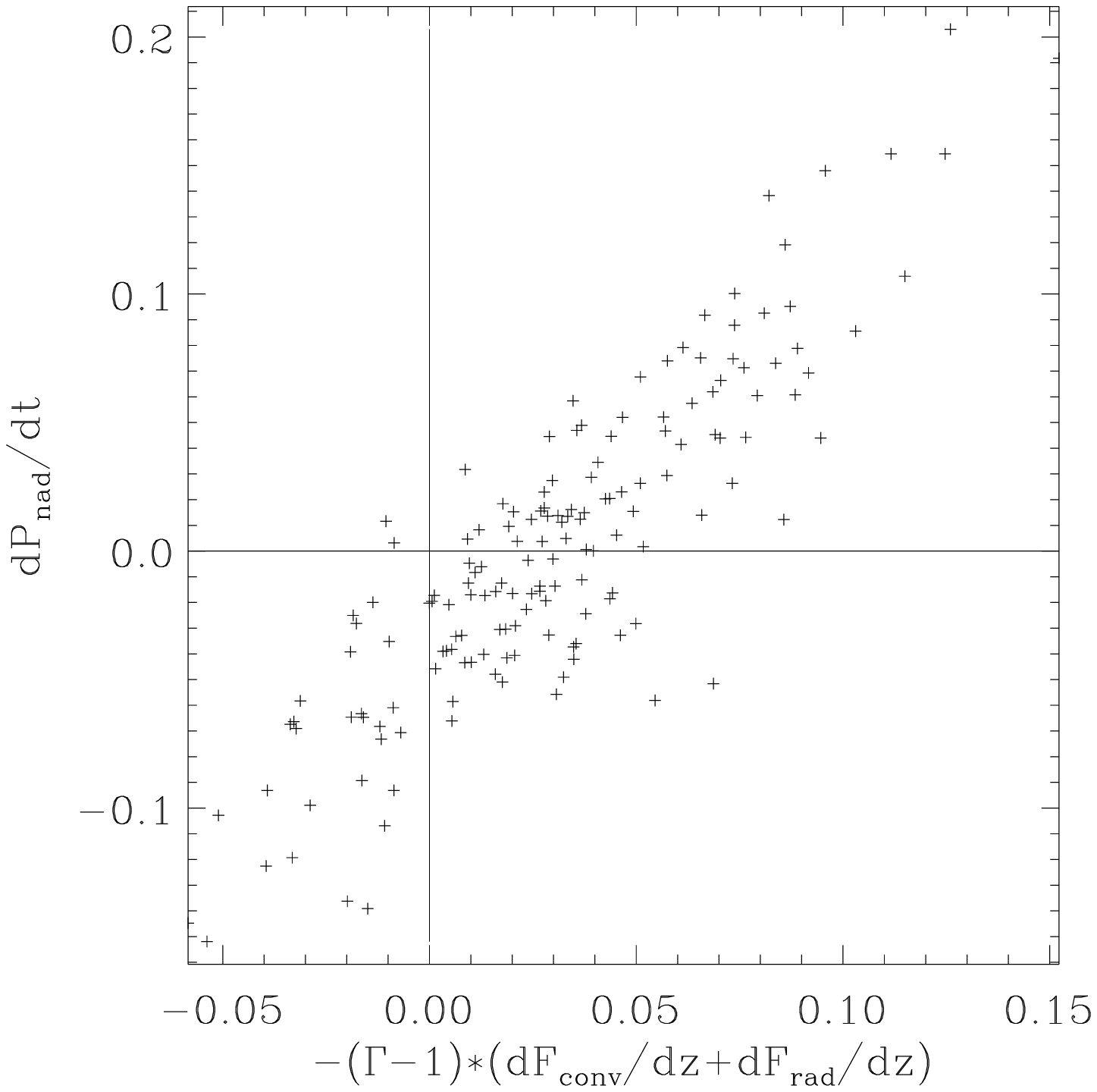}{8.0cm}{
   Correlation of the divergence of the convective and radiative
   fluxes, at z=100 km, multiplied by ($\Gamma_3-1$) with the time
   derivative of the non-adiabatic pressure at the surface.  The units
   are 10$^3$ ergs/cm$^3$/s.  The close correlation shows that the
   slight instantaneous imbalance in the radiative and convective flux
   divergences is the primary source of non-adiabatic pressure
   fluctuations.  
}

Where in space does this driving occur?  The warm granules have only
small non-adiabatic pressure fluctuations.  Large, negative
fluctuations are concentrated in the downdrafts
(Figs.~\ref{figure13}, \ref{figure14}).  

The maximum mode driving occurs in the frequency range of 3 -- 4 mHz,
by non-adiabatic pressure fluctuations in the same frequency range.  By
filtering the time sequence of these fluctuations we see 
(Fig~\ref{figure15}) that in the peak driving range 
also the
driving occurs predominantly in the intergranule lanes and near the
edges of granules.  The high frequency power near granule edges is
due to the motion of the granule boundaries over the one hour time
interval on which the filtering was performed.  This is, in part, a result 
of changes in granule size as they evolve.
No direct correlation of non-adiabatic pressure fluctuations in the
range of 2-5 mHz with velocity is seen.  Keep in mind, however, that a
correlation plot does not reveal correlations of events that happen in
the neighborhood of one another.

%\FIG{pnad_vz_corr-4mHz-z=0}{8.0cm}{
%   Correlation of non-adiabatic pressure fluctuations in the 2.5 -- 5 mHz
%   range with vertical velocity at the surface.  In this frequency range,
%   where the driving is maximal, the largest $\delta P_{\rm nad}$  occur
%   inside and at the edges of the intergranular lanes.
%}

\subsection{Excitation Source}
\FIG{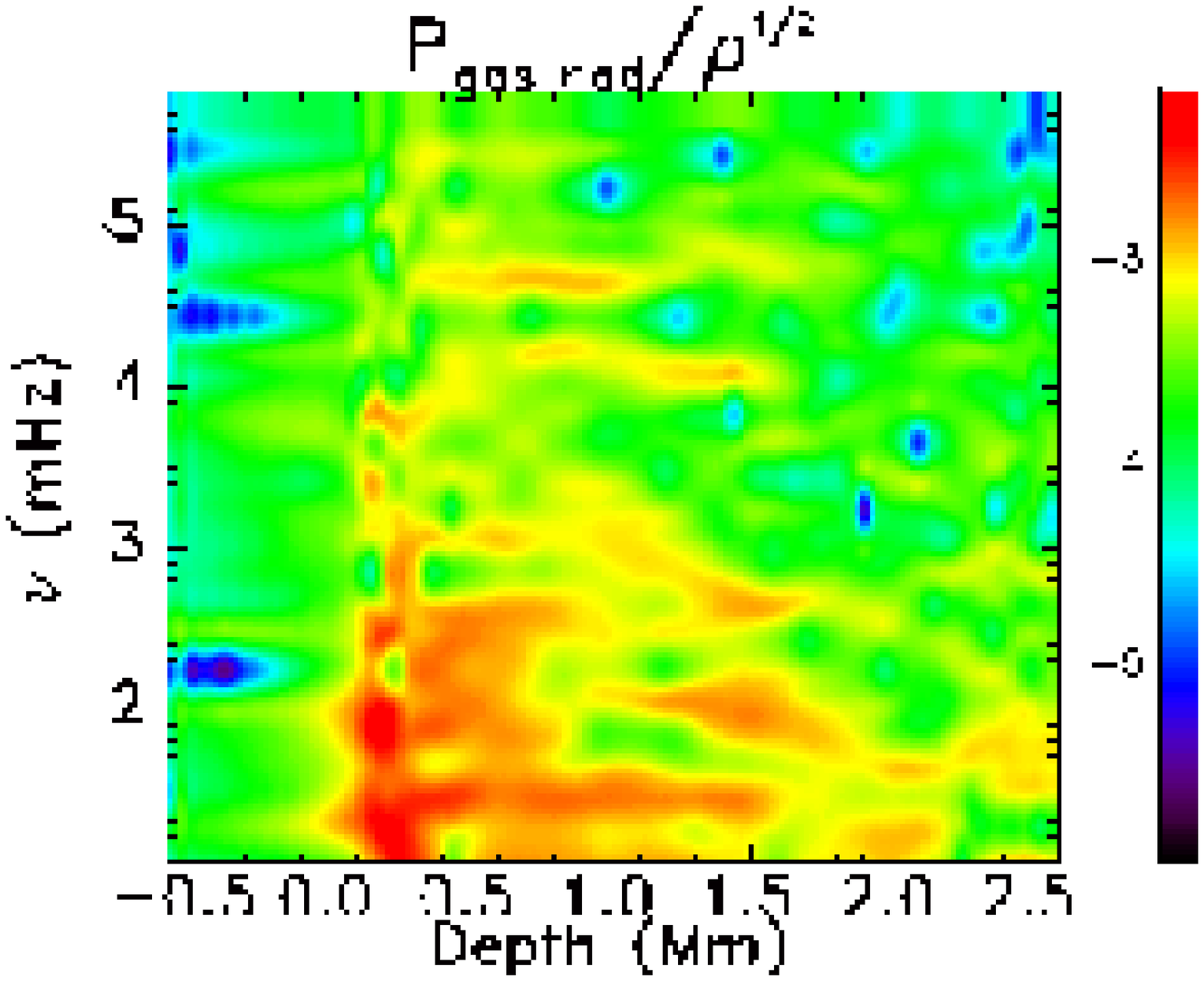}{7.0cm}{
Logarithm of the gas pressure fluctuations scaled by the square root of
the density, as a function of depth and frequency.  The color scale is 
identical to the following image of the turbulent pressure fluctuations.  
With increasing frequency the gas pressure fluctuations become more rapidly
concentrated near the surface and decrease more rapidly in strength
than the turbulent pressure fluctuations.
}
\FIG{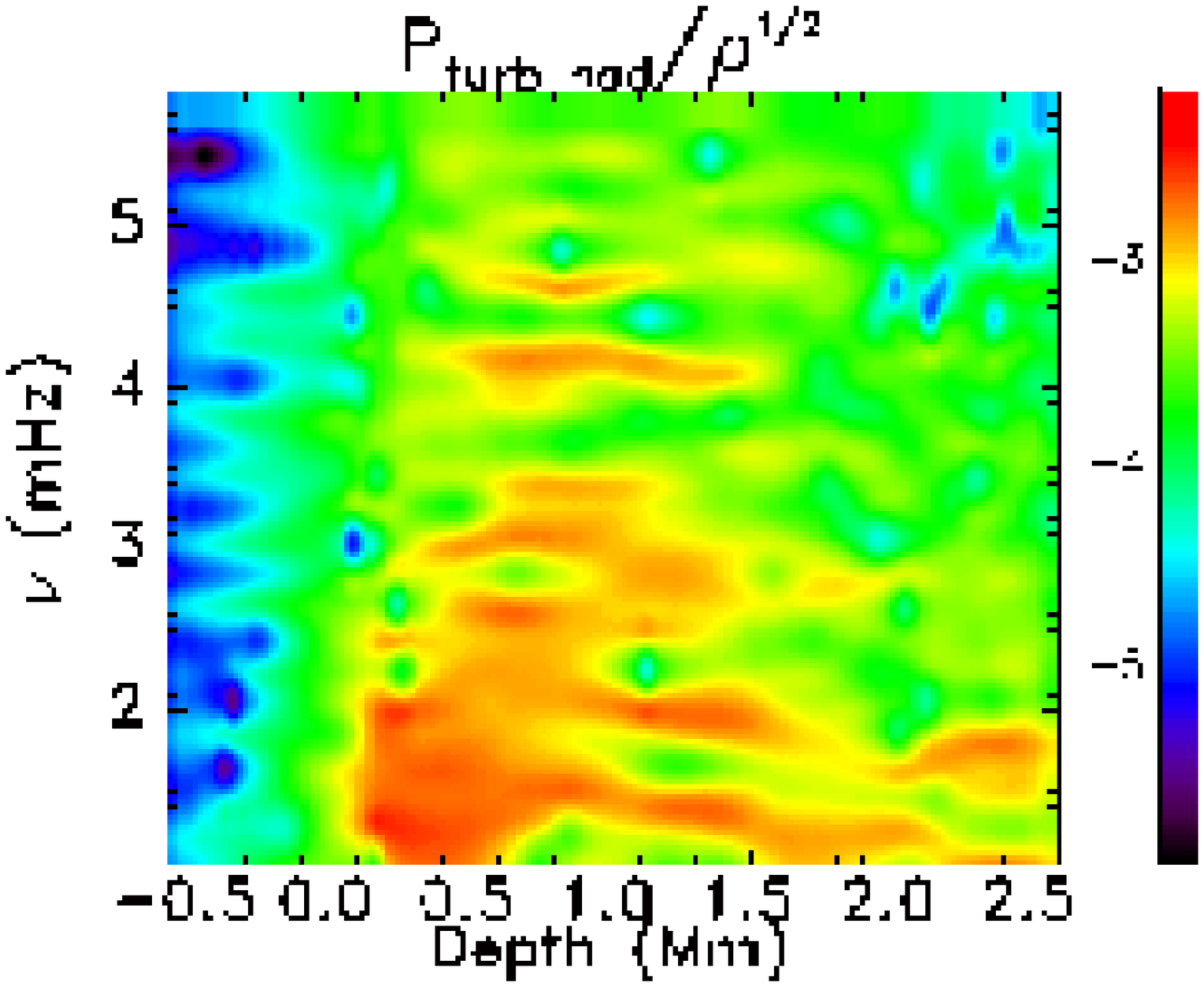}{7.0cm}{
Logarithm of the turbulent pressure fluctuations scaled by the square
root of the density, as a function of depth and frequency.  The color scale 
is identical to the preceding image of the gas pressure fluctuations.  The
turbulent pressure fluctuations extend deeper and decrease less rapidly
in magnitude with increasing frequency than the gas pressure
fluctuations.  As a result, turbulent pressure is the primary source of
$p$-mode excitation in the peak driving range of 3-4 mHz.
}
\FIG{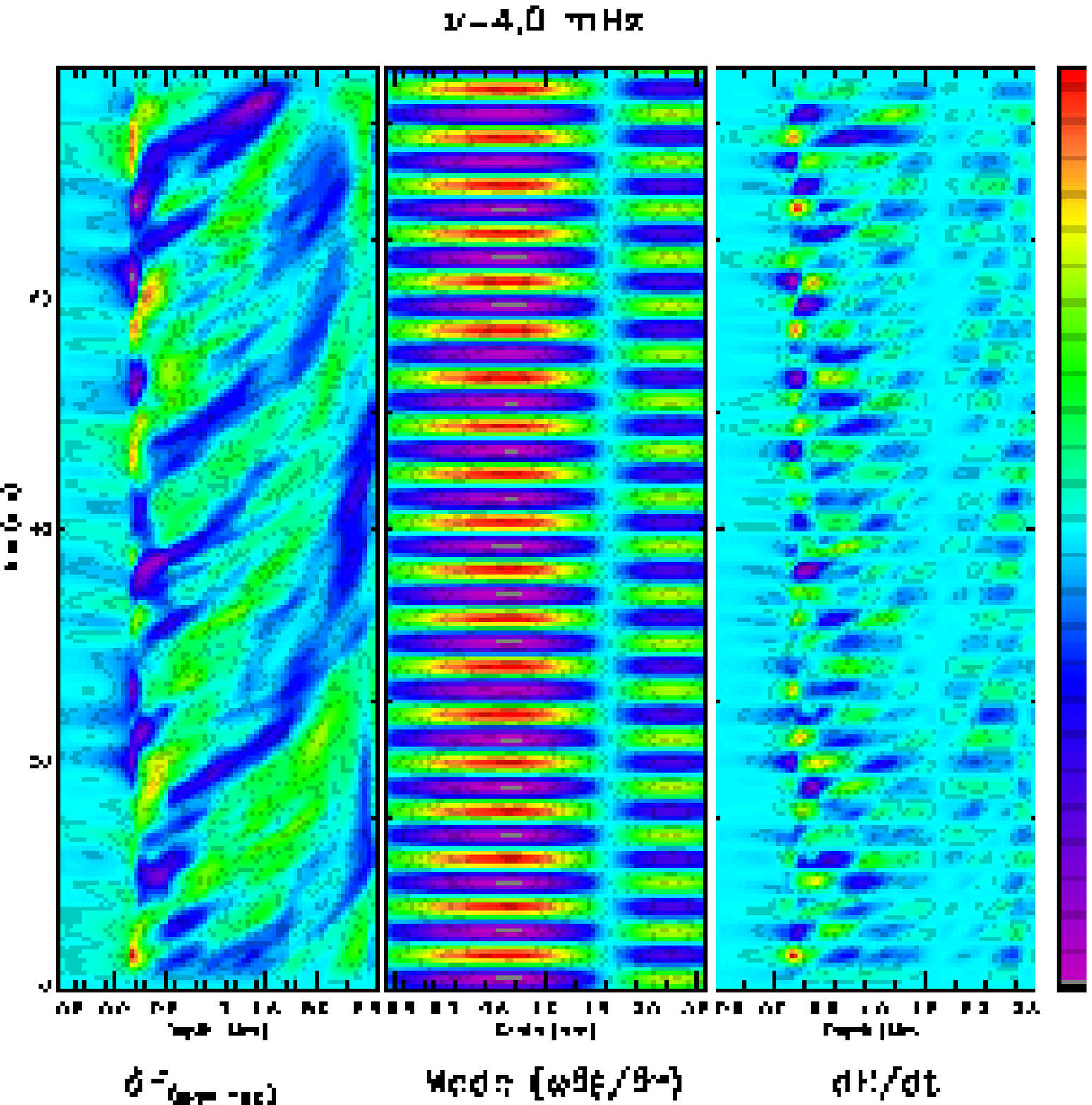}{7.0cm}{
   Integrand of the stochastic work integral for non-adiabatic gas
   pressure only, at 4 mHz, as a function of depth and time (right
   panel, eqn~\ref{eq_edot}) and the two terms that contribute to it:
   the horizontally averaged, non-adiabatic gas pressure fluctuations,
   multiplied by $\langle \rho \rangle^{-1/2}$ (left panel) and the
   coherent mode density fluctuations multiplied by $\langle \rho
   \rangle^{1/2}$ (center panel).  The non-adiabatic gas pressure
   fluctuations (left panel) are produced by the imbalance of the
   convective and radiative energy transport and are largest close to
   the surface.  The color scale goes from maximum negative to maximum
   positive value for each variable. 
}
\FIG{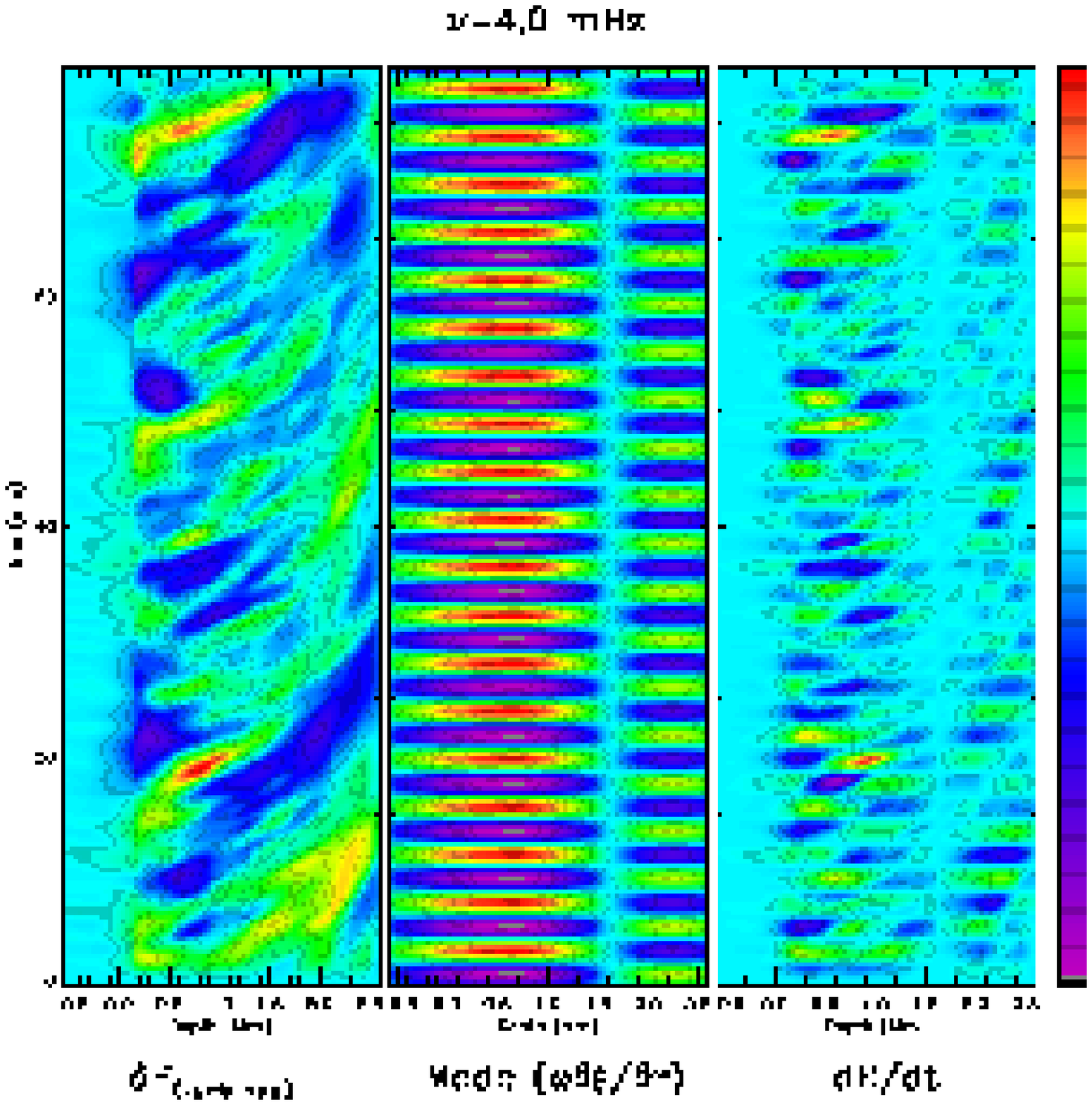}{7.0cm}{ 
   The same as figure \ref{figure23}, but for the
   non-adiabatic turbulent pressure only.  The turbulent pressure
   fluctuations are more coherent in depth and time than the 
   non-adiabatic gas pressure fluctuations so they produce a more 
   coherent contribution to the work integral.  
}
\FIG{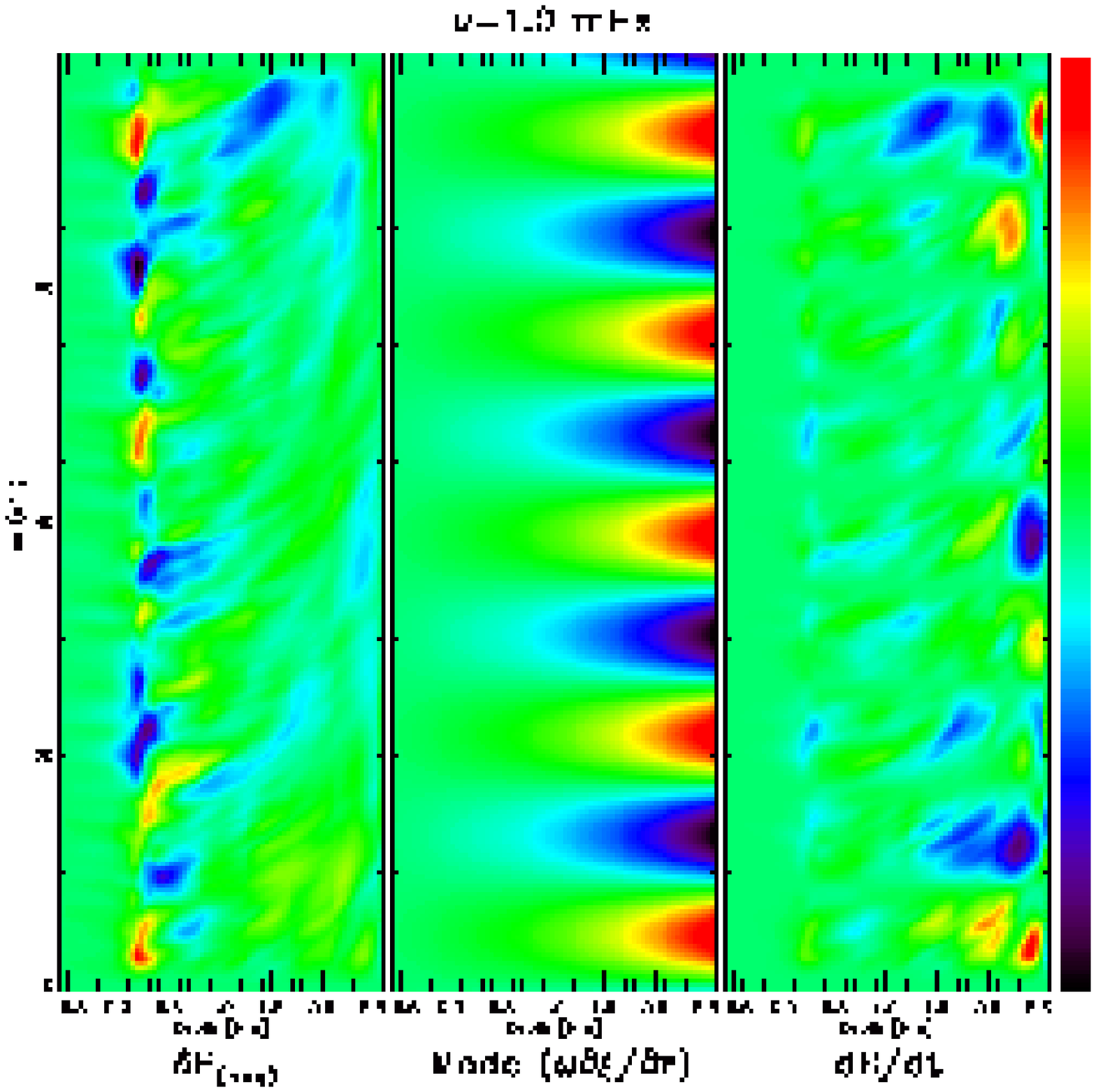}{7.0cm}{
   Integrand of the stochastic work integral as a function of depth and
   time (right panel, eqn~\ref{eq_edot}) and the two terms that
   contribute to it: the horizontally averaged, non-adiabatic pressure
   fluctuations, multiplied by $\langle \rho \rangle^{-1/2}$ 
   (left panel) and the coherent mode density fluctuations multiplied
   by $\langle \rho \rangle^{1/2}$
   (center panel).  The non-adiabatic pressure fluctuations (left
   panel) are produced by the imbalance of the convective and radiative
   energy transport and are largest close to the surface.  The color scale
   goes from maximum negative to maximum positive value for each variable.
}
\FIG{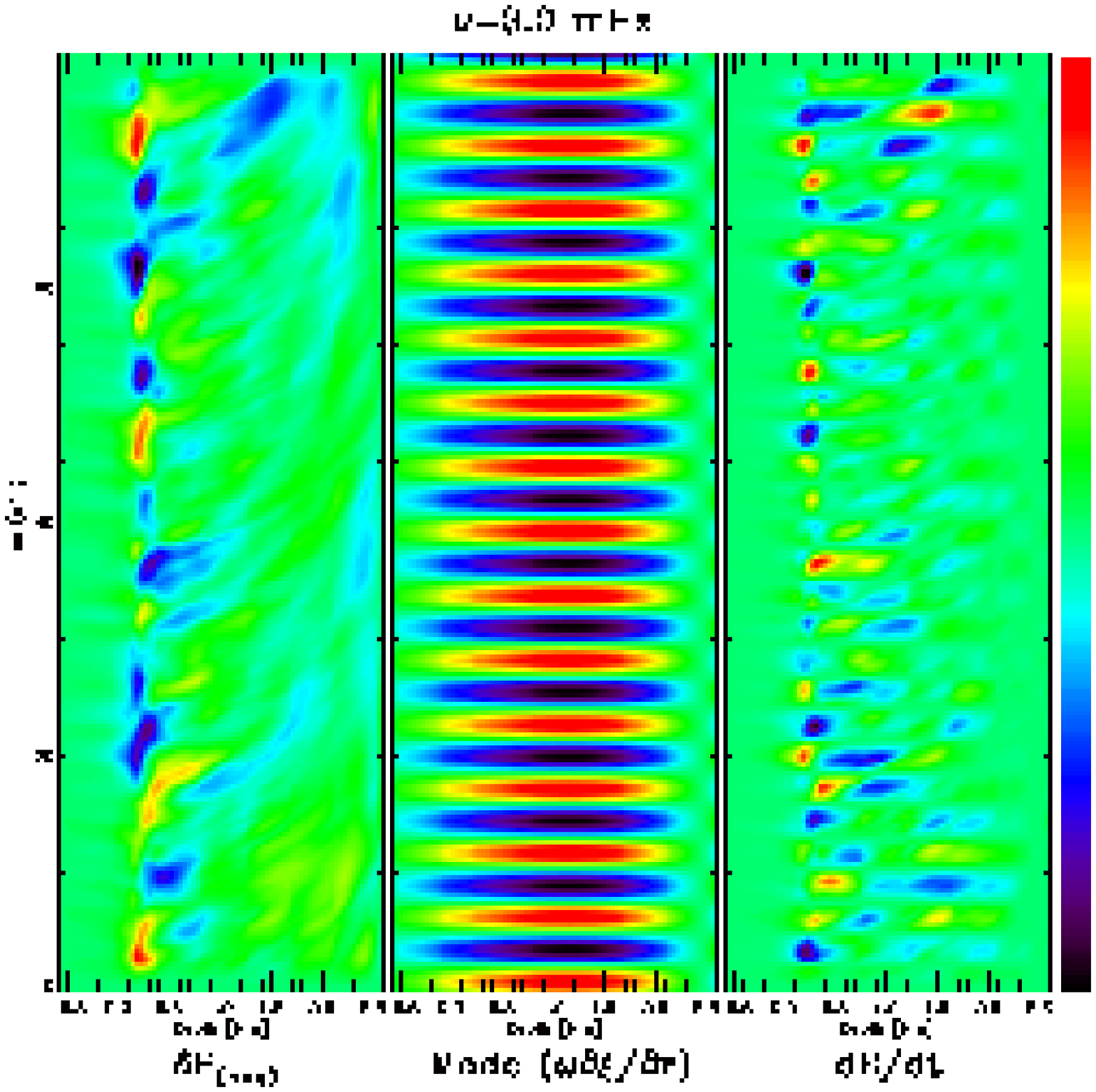}{7.0cm}{Same as Fig~\ref{figure25}, but for $\nu=3$
   mHz.  At higher frequencies the mode compression (center panel)
   and hence the work (right panel)
   becomes more concentrated toward the surface.
}
\FIG{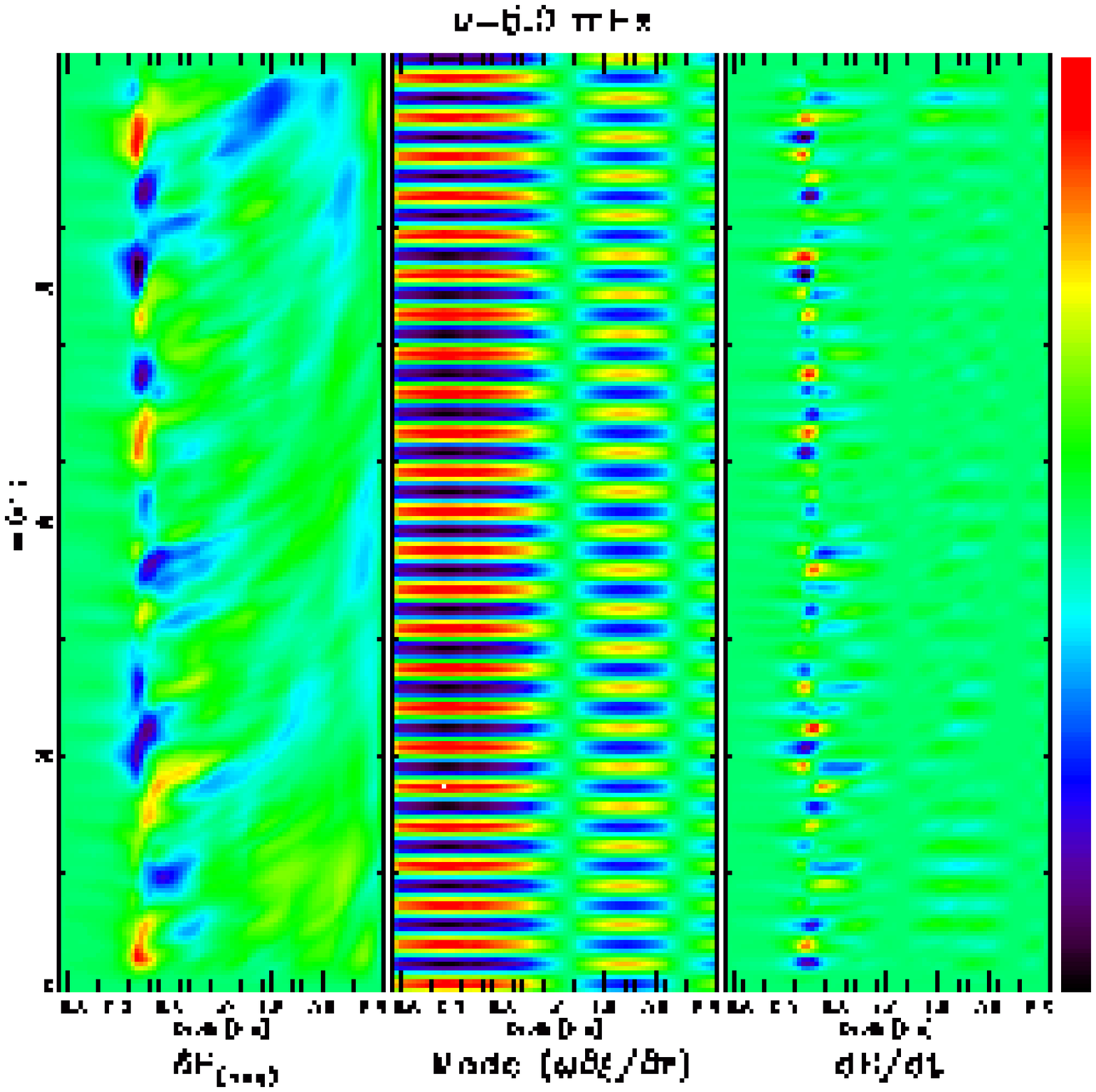}{7.0cm}{Same as Fig~\ref{figure25}, but for $\nu=5$
   mHz. The compression and work are even more concentrated toward the 
   surface and the mode has several radial nodes.}

What is the source of the non-adiabatic pressure fluctuations?  Is it 
entropy fluctuations or Reynolds stresses?  Both play a role, but the
primary source of mode driving is turbulent pressure fluctuations
(Reynolds stresses)
(Fig~\ref{figure16}).  This is surprising, since
the non-adiabatic gas pressure power is larger than the turbulent
pressure power near the surface (Fig~\ref{figure9}).
The non-adiabatic gas and turbulent pressures contribute comparably to 
the work near the surface, but the contribution of the
turbulent pressure extends deeper and provides the dominant
contribution to the total work (Fig~\ref{figure17}).

The gas pressure fluctuations are significantly larger than the
turbulent pressure and have maxima at the frequencies of the three
radial modes of the simulation.  The non-adiabatic pressure fluctuation
power, however, varies smoothly across these frequencies indicating
that it is primarily due to stochastic convective processes
(Fig~\ref{figure9}).  

There 
is a tight correlation between the non-adiabatic pressure fluctuations
at the surface and the emergent intensity (Fig~\ref{figure18}),  
which indicates that it is the fluctuating cooling at the surface that 
is the main source of stochastic mode excitation there.  Indeed, the source 
of entropy fluctuations is the cooling of fluid that approaches 
optical depth unity \citep{Stein+Nordlund98gran}.  
This correlated noise is believed to be
responsible for the difference in asymmetry of the modal power spectra
observed in velocity and intensity \citep{Nigam+98,Kumar+Basu99a}.
Our discussion in terms of
non-adiabatic pressure fluctuations is equivalent to the discussion in
terms of entropy fluctuations by \citet{Goldreich+94b}.

We can use the energy equation to determine the 
processes producing the non-adiabatic gas pressure fluctuations 
\citep{Stein+Nordlund91osc}.
\begin{equation}
\delta P_{\rm nad} = \delta P -{{\Gamma_{1} P} \over \rho} \delta 
\rho = \left( {{\partial P} \over {\partial s}}\right)_{\rho} \delta s = 
(\Gamma_{3} - 1) \delta Q
\ ,
\end{equation}
and
\begin{equation}
{{D Q} \over {D t}} = - {\partial \over {\partial z}} \left( F_{\rm 
rad} + F_{\rm conv} \right) + \breve{\mathbf u} \cdot \nabla P + 
Q_{\rm diss}
\ ,
\end{equation}
where $\breve{\mathbf u} = {\mathbf u} - 
\langle\rho {\mathbf u}\rangle/\langle\rho\rangle$ is 
the convective component of the velocity.  The divergence of the 
radiative and convective fluxes are the dominant terms, but they 
nearly cancel each other since energy transport shifts from convective 
to radiative near the surface.  It is their slight instantaneous 
imbalance locally that leads to the non-adiabatic pressure fluctuations
(Figs~\ref{figure19} and \ref{figure20}).

The separate contributions of the gas and turbulent pressure to the total
non-adiabatic pressure spectrum are shown in figs. \ref{figure21}
and \ref{figure22}, with the same color scale in both.  Near the
surface, there is clearly more power in the gas pressure.  However, in
the peak driving range of 3-4 mHz, there is more power in the
non-adiabatic turbulent pressure below the surface.  

Another reason for the dominance of turbulent pressure in the mode
driving is revealed in figs. \ref{figure23} and
\ref{figure24}.  The left panels in each show the
non-adiabatic gas and turbulent pressures (scaled by $\rho^{-1/2}$) as
a function of time and depth.  The turbulent pressure varies more
slowly with depth and has a longer time scale than the gas pressure.
The middle panels show one realization of the mode density or
compression (scaled by $\rho^{1/2}$).  The right panels show the work
integrand, $\delta P_{{\rm nad} \omega} \partial \xi_{\omega} / 
\partial z E_{\omega}^{-1/2}$.  The gas
pressure contribution to the integrand is more concentrated at the
surface and alternates sign with time and depth more rapidly than the
turbulent pressure contribution.  The greater coherence of the turbulent
pressure allows it to make a greater contribution to the net work.

The very largest pressure perturbations are associated with the sudden 
initiation of downdrafts and produce waves that propagate up into 
the chromosphere and steepen into shocks 
\citep{Skartlien98thesis,Skartlien+99waves}.  
These events may correspond to the large individual 
acoustic events observed by \citet{Rimmele+95} and \citet{Goode+98}. 
They may be the tail of the distribution of the stochastic driving 
process.  A comparison of observed acoustic events and those in the 
simulation has been made by \citet{Goode+98}.

\section{Summary}

How does this all fit together?  Figs~\ref{figure25} --
\ref{figure27} show the time variation of the horizontally averaged
non-adiabatic pressure fluctuations, the density (compression) in the
modes and the product of the two which is the integrand of the work
integral (eqn~\ref{eq_edot}) as a function of depth and time for modes
with frequencies 1,3, and 5 mHz.

The dominant contribution to the work comes from the turbulent pressure 
(Reynolds Stress).  Near the surface non-adiabatic gas pressure 
fluctuations, produced by an instantaneous imbalance between the 
divergences of the radiative and convective fluxes, also contribute.  
Their divergence is individually large, but since
energy transport switches between convection and radiation in the
surface layers, they are nearly equal and opposite.  At each instant
they do not exactly cancel, which leads to heating and cooling and hence
entropy fluctuations.  
Excitation is small at low frequencies due to
mode properties -- the mode compression decreases and the mode mass
increases at low frequency.  In addition, at low frequency the mode
amplitude is small near the surface where the non-adiabatic pressure
fluctuations are large.  (At very low frequencies, driving occurs 
deeper than the simulation domain.)  
Excitation is small at high frequencies due to
the pressure fluctuation spectrum -- pressure fluctuations become small
at high frequencies because they are due to convective motions which have 
a longer time scale than the dominant $p$-mode periods.
Large non-adiabatic pressure fluctuations occur primarily in the
intergranular lanes (due to turbulence and surface radiative cooling) 
and near the edges of granules (due to granule expansion and collapse).

\section*{Acknowledgments}

We thank Dali Georgobiani for making the 43 hour simulation.
This work was supported in part by NASA grants NAG 5-4031 and NAG
5-8053, NSF grants AST 9521785 and AST 9819799, and the Danish Research
Foundation, through its establishment of the Theoretical Astrophysics
Center.  The calculations were performed at the National Center for
Supercomputer Applications, which is supported by the National Science
Foundation, at Michigan State University and at UNI$\bullet$C,
Denmark.  This valuable support is greatly appreciated.

%%%%%%%%%%%%%%%%%%%%%%%%%%%%%%%%%%%%%%%%%%%%%%%%%%%%%%%%%%%%%%%%%%%%
%\clearpage
\bibliographystyle{apjbib}
\bibliography{aajour,bob,convection,aake,oscillations,waves,books}
%%%%%%%%%%%%%%%%%%%%%%%%%%%%%%%%%%%%%%%%%%%%%%%%%%%%%%%%%%%%%%%%%%%%

\renewcommand{\textheight}{22cm}

\appendix
\section{Evaluation of the Mode Excitation Rate
(eqn~\ref{eq_edot})}

The mode excitation rate (eqn~\ref{eq_edot})
can be evaluated in two different ways: in Fourier
space and in real space.  We describe both methods.  The variables
that appear in eqn \ref{eq_edot} for the mode excitation are the
non-adiabatic total pressure fluctuation, the derivative of the mode
displacement and the mode energy.  These quantities are calculated as
follows for both methods of evaluating the excitation.

The non-adiabatic pressure fluctuations are obtained from the
simulation results.  First the gas pressure, turbulent pressure
($P_{t}= \langle\rho u_{z}^{2}\rangle$) and density are averaged over
horizontal planes and saved at 10 s or 30 s intervals.  These are then
interpolated to the Lagrangian frame, that moves with the vertical
(radial oscillation) motions.  The Lagrangian frame is determined by
calculating the mass column density at each time and interpolating
the variables to the time average mass column density.  Next, the
non-adiabatic total (gas plus turbulent) pressure is calculated from
\EQ
P^{\rm nad} = \left( \ln (P_{\rm gas}+P_{\rm turb}) - \Gamma_1 \ln \rho 
\right) (P_{\rm gas}+P_{\rm turb}) \ .
\label{eqn_pnad}
\EN
Finally, the fluctuation of the non-adiabatic total pressure about 
its time average, 
\EQ
\delta P^{\rm nad}(z,t)=P^{\rm nad}(z,t)- \langle 
P^{\rm nad}(z,t)\rangle_t \ ,
\label{eqn_pnad_fluct}
\EN
is determined.

The mode displacement $\xi_{\nu}(z)$ for the radial mode of frequency
$\nu$ is obtained from the eigenmode calculations of
Christensen-Dalsgaard, using his spherically symmetric model S
\citep{JCD+96science}.  
Alternatively, one could extract modes directly from the numerical
simulations, using Fourier decomposition.  A drawback with this
method is, apart from slightly noisier mode structure, that there are
only three resonant modes in the shallow simulation box---the 35 radial
modes from standard solar models provide much better frequency coverage.
As we have shown (fig~\ref{figure2}), the modes from the simulation 
are essentially the
same as the solar modes at the resonant frequencies of the computational 
domain, where the solar modes have a node at the depth of the bottom of the
computational domain.

The mode energy is calculated from the displacement according to 
eqn.~\ref{eq_enorm}.
The mode displacements are interpolated to the simulation grid at the 
same total pressure as in the Christensen-Dalsgaard model and the 
derivative of the displacement is calculated.  Since the derivative 
of the mode displacement always appears normalized by the square root 
of the mode energy, the mode amplitude normalization cancels.

% Fourier space method

In Fourier space, the work integral which appears in the energy input
rate to the modes (eqn~\ref{eq_edot}) is evaluated by taking the time
Fourier transform of the total non-adiabatic pressure fluctuations at 
each depth, $\delta P^{\rm nad}(z,t)$.  
For each frequency and depth, this is multiplied by the
spatial derivative of the mode displacement at
that depth interpolated to the frequencies of the Fourier transform
(and normalized by the square root of the mode energy), $\omega (\partial 
\xi_{\omega} / \partial r) E_{\omega}^{-1/2}$.  The spatial dependence of
the modes varies slowly and continuously with frequency, so such 
interpolation is possible.  This product is integrated over the depth
of the simulation domain for each frequency.  The energy input rate to
the modes is the square of the absolute value of this work integral,
divided by the frequency interval for the Fourier transform, $\Delta
\nu$, (which equals multiplying by the time interval of the simulation), 
multiplied by the area of the simulation (36 ${\rm Mm}^2$) and
divided by eight (Eqn~\ref{eq_edot}).

% Real space method
In real space, the integrand of the work integral for each 
frequency, $\nu$, is calculated as the 
non-adiabatic total pressure fluctuation at each depth and time,
$\delta P^{\rm nad}(z,t)$, multiplied by the normalized derivative of
the mode displacement for that frequency at each depth,
$\omega ({\partial \xi_{\omega}(z)} / {\partial z}) / E_{\omega}^{1/2}$,
multiplied by a phase function $\cos \left(\phi+\omega t\right)$
for each time, $t$, a snapshot was saved in the simulation (10 s or 30 s
intervals).  For each depth, this is summed over all saved snapshots
in the longest interval that is an integral number of mode periods
for the given mode frequency $\nu$, and divided by the number of
snapshots summed over to get the average value.  This integrand is
then integrated over depth.  This is done for two values of the
phase, $\phi = 0 \ {\rm and}\  \pi/2$ (since the phases between the modes
and the pressure fluctuations are random, we average these two
orthogonal cases).  The value of the integral for each phase is
squared and these two values are summed.  The energy input rate 
to the mode at this frequency is this average multiplied 
by the time interval integrated over (for that frequency), 
multiplied by the area of the simulation (36 ${\rm Mm}^2$), 
and divided by eight.

\end{document}